\documentclass[preprint]{aastex631}
\usepackage{amsmath, amsfonts, amssymb}
\usepackage[normalem]{ulem}
\hypersetup{linkcolor=magenta,citecolor=blue,filecolor=cyan,urlcolor=blue}

\received{January 22, 2025}
\revised{Month Day, Year}
\accepted{Month Day, Year}
\submitjournal{ApJ}

\begin{document}

\title{The W Transform: an additional powerful tool for astrophysical time-series analysis}

\newcommand{\paulo}[1]{\textcolor{blue}{#1}}


%
%
\author[0000-0002-7353-536X]{P. D. S. de Lima}
\affiliation{Departamento de F\'isica Te\'orica e Experimental, Universidade Federal do Rio Grande do Norte, Campus Universit\'ario, Natal, RN, 59072-970, Brazil}
\affiliation{School of Physics, Trinity College Dublin, Dublin, Ireland} 
\author[0000-0002-3916-6441]{D. O. Fontinele}
\affiliation{Departamento de F\'isica Te\'orica e Experimental, Universidade Federal do Rio Grande do Norte, Campus Universit\'ario, Natal, RN, 59072-970, Brazil}
\author[0000-0002-2023-7641]{R. L. Gomes}
\affiliation{Departamento de F\'isica Te\'orica e Experimental, Universidade Federal do Rio Grande do Norte, Campus Universit\'ario, Natal, RN, 59072-970, Brazil}
\author[0000-0001-5578-7400]{B. L. Canto Martins}
\affiliation{Departamento de F\'isica Te\'orica e Experimental, Universidade Federal do Rio Grande do Norte, Campus Universit\'ario, Natal, RN, 59072-970, Brazil}
\author[0000-0001-8462-4280]{J. M. de Ara\'ujo}
\affiliation{Departamento de F\'isica Te\'orica e Experimental, Universidade Federal do Rio Grande do Norte, Campus Universit\'ario, Natal, RN, 59072-970, Brazil}
\author[0000-0002-1044-4348]{Y. Wang}
\affiliation{Imperial College London, Resource Geophysics Academy, London SW72BP, UK}
\author[0000-0001-8218-1586]{J. R. De Medeiros}
\affiliation{Departamento de F\'isica Te\'orica e Experimental, Universidade Federal do Rio Grande do Norte, Campus Universit\'ario, Natal, RN, 59072-970, Brazil}
\author[0000-0001-5845-947X]{I. C. Le\~ao}
\affiliation{Departamento de F\'isica Te\'orica e Experimental, Universidade Federal do Rio Grande do Norte, Campus Universit\'ario, Natal, RN, 59072-970, Brazil}

\correspondingauthor{I. C. Le\~ao}
\email{izan@fisica.ufrn.br}

\begin{abstract}

This work introduces the W Transform (WT) for astrophysical applications and explores its similarities and differences to the Continuous Wavelet Transform (CWT), in photometric stellar light curves (LCs). We tested both techniques on synthetic data, producing LCs with different noise levels and numbers of gaps, and applied them to LCs from TESS space mission. The analysis focused on six stars: the solar analog TIC 276293046, the ultra-cool dwarf TIC 359892714, the $\gamma$ Dor pulsator TIC 164173105, the eclipsing binary TIC 355151781, the hosting planet star TIC 142276270, and the brown dwarf TIC 102076870. Overall, the results were in agreement, except for stars with LCs presenting sector-to-sector differences in structures or changes in the number of gaps, reflecting time-dynamic effects in the same target, which affect how time-dependent variability features appear in the time–frequency representation. For such LCs, the WT spectra are more stable and accurate in recovering signal amplitudes than the CWT maps, which exhibit distortions and limitations, showing that the inherent flexibility of the WT method in adjusting the dominant frequency could enhance time resolution even in LCs with complex or variable structures. We conclude that the WT is a robust tool for the analysis of stellar LCs, and that both methods may complement each other in an in-depth analysis of a periodic astrophysical signal. The present results suggest that the WT is a robust and accurate tool for analyzing stellar LCs, expanding the range of numerical techniques available for analyzing different stellar periodic phenomena from photometric observations.

\end{abstract}



\section{Introduction} \label{sec:intro}

Thanks to the advent of ultra-high-precision photometry from space missions, science has witnessed, over the past 25 years, fundamental breakthroughs in advancing our understanding of the atmosphere and interior of stars. From samples of photometric light curves (LCs) counted by the dozens over the years up to 2010, space missions brought LCs to tens of thousands of stars with the CoRoT \citep{2009IAUS..253...71B} and Kepler \citep{2010Sci...327..977B} missions, and to tens of millions with the TESS mission\footnote{The Transiting Exoplanet Survey Satellite. \url{https://tess.mit.edu/}.} \citep{2015JATIS...1a4003R}. The growing amount of data requires considerable effort for treatment, with various algorithms used to extract information on different physical stellar phenomena that shape the light curves. This includes autocorrelation functions \citep[e.g.,][]{2014ApJS..211...24M}, discrete and fast Fourier transforms \citep[e.g.,][]{cooley1965fft}, Lomb-Scargle periodograms \citep{1976Ap&SS..39..447L, 1982ApJ...263..835S}, the Hilbert-Huang transform \citep{Huang1998hhtransform}, and the Wavelet transform \citep{Torrence1998}. The TESS space mission was launched in 2018 and features an observational strategy that enables the mapping of an entire hemisphere in a single year. Most of the sky is monitored with 27 days of continuous observations, while a small area around the poles receives more or less continuous coverage throughout the entire year. TESS completed its 6-year, all-sky nominal survey in July 2024, dramatically increasing the number of stars with high-quality LCs. Therefore, TESS is enabling the study of a variety of stellar phenomena on timescales that have never been achieved before. 

Stellar light curves can be decomposed into several frequencies represented in the power spectrum, allowing us to determine the periodic components of the data that may be related to different physical phenomena, including rotation modulation, pulsation, flares, and planetary transits. From the application of the Fourier transform to the signal, we can obtain its frequency–amplitude representation. Nevertheless, since the stellar light curves present events that do not occur periodically, such as the growth and decay of an active region, one of the main focuses, in addition to obtaining the spectral composition, which offers an idea of rotation behavior and pulsation modes, is to follow the time–frequency behavior of those events and identify any specific signature to a particular stellar variability even if the light curve presents some noise or singularities. As shown by \citet{bravo_2014}, the time localization of the spectral components is fundamental to a more complete analysis of LC. The application of the wavelet transform to the signal, producing its time-frequency portrait, is a determinant step, including the definition of the persistence of a physical phenomenon acting on the LC. In fact, several studies have applied the Classical Wavelet Transform (CWT) in the analysis of stellar photometric light curves \citep[e.g.,][]{fontinele2024,gomes2024,Santos2021,bravo_2014,2019MNRAS.484.3935D}. The literature also reports the studies by \citet{Ramon-Ballesta2022}, who applied the synchrosqueezed wavelet transform \citep{Daubechies2011} to perform a time-frequency analysis of the light curves of two $\delta$ Scuti stars, and by \citet{Dorn-Wallenstein2020}, who applied the wavelet Z-transform \citep{Foster1996} to analyze the short-term variability of evolved massive stars. In addition to identifying periodicities and specific variability signatures, such as rotation and pulsation, the studies by \citet{fontinele2024}, \citet{gomes2024}, \citet{bravo_2014}, \citet{2019MNRAS.484.3935D}, and \citet{Dorn-Wallenstein2020} also aimed to quantify the persistence level of these specific variability signatures.

The present study provides a detailed analysis of LCs collected by TESS, featuring signatures of various stellar physical phenomena, including rotation and pulsation, utilizing the W Transform (WT) \citep{wtpaper, wang2022time, wang2024}. This transform has some relevant features, including the fact that the spectral energy is concentrated around the dominant frequency, the implementation is numerically stable as it avoids any potential frequency singularity, and the Gaussian window function is defined using a nonstationary frequency weight. The WT has been successfully applied in the treatment of various geophysical problems. In general, it improves the time resolution of the low-frequency spectrum in the detection of seismic anomalies and abnormal features in near-surface geophysics \citep[e.g.,][]{wtpaper, tp_wt, li2022, luo2023, wu2024}. One key motivation in the present work is to enlarge the horizon of numerical procedures for the analysis of photometric stellar LCs, making it possible to reveal additional features present in the LCs. For this purpose, we have selected a sample of six stars observed by the TESS space mission, which offers unique high-quality photometric data. The paper is organized as follows. Section~\ref{transform} presents the theoretical framework of the~WT. Section~\ref{comparison} shows the initial results derived from synthetic data, while Section~\ref{results} provides the main results based on TESS~data. Finally, a summary is presented in Section~\ref{sec:conclusion}.

\section{The W Transform} \label{transform}

Time-frequency analysis (TFA) extracts nonstationary features of time series data to study how the signals present in a time series behave over time. Since time-frequency mapping is not a unique process, different TFA methods have been proposed in the literature. Most methods compute the classical Fourier transform within different time windows along a time series by convolving the time series with a given kernel. Among these are included the pioneering Gabor transform (GT)~\citep{gabor, cohen}, the continuous wavelet transform (CWT)~\citep{cwt}, and the S transform (ST)~\citep{st, mansinha1, mansinha2, symon}, to name but a few. The resolution in the time-frequency domain for each of these methods depends on how the window function is used to convolve the signal, that is, how the local kernel is defined. In this regard, the GT fixes the window size, leading to poor time-frequency resolution. On the other hand, the window size can be varied in the ST and the CWT, providing a better multi-resolution analysis tool. The latter transforms are indeed closely related~\citep{ventosa}.

The adaptive time-frequency resolution in the CWT is obtained by considering the window function in terms of daughter wavelets 
\begin{equation}
    \psi_a(t) = \frac{1}{\sqrt{a}}\psi\left(\frac{t}{a}\right)\,,\label{eq:d_wavelet}
\end{equation}
which are time-localized zero average functions constructed from a mother wavelet $\psi(t)$ and deformed by a dilatation parameter $a>0$. Depending on the value of $a$, the wavelet is dilated (contracted) to improve the frequency (time) resolution at the price of the time (frequency) resolution~\citep{mallat}. Moreover, the factor $\sqrt{a}$ in Equation \eqref{eq:d_wavelet} ensures that the energy is conserved between wavelets with different dilation parameters. Therefore, the CWT for a given signal $s(t) \in L^2(\mathbb{R})$ is defined by
\begin{equation}
	\text{CWT}(t, a) = \frac{1}{\sqrt{a}}\int_{-\infty}^{\infty}s(\tau)\psi^\star_a(\tau - t)\,d\tau\,, \label{eq:cwt}
\end{equation}
where the symbol $\star$ denotes complex conjugation. Additionally, the spectrograms computed from Equation \eqref{eq:cwt} can be converted to the time-frequency domain using the pseudo-frequency representation~\citep{pseudo_freq}, and here we use a 2nd-order Morlet wavelet. Further details concerning the analysis of stellar light curves using the CWT can be found in~\cite{bravo_2014}.

Despite the success of the CWT in the detection of astrophysical events\textcolor{red}{~\citep[e.g.,][]{Ghosh2023, Gurgenashvili2022, 2020ApJS..250...20C, 2019MNRAS.484.3935D, bravo_2014}}, its construction has some limitations. For example, as the time resolution decreases at low-frequency bands, it may suppress specific stellar events in LCs. To overcome this limitation, the recent WT~\citep{wtpaper} incorporates the dominant frequency information by generating a time-varying window function.  This modification enhances the time resolution in both low and high frequencies by shifting the spectral energy towards the instantaneous frequency of the signal at each instant of time. In this regard, the WT is then defined as~\citep{wtpaper, wang2022time}
\begin{equation}
	\mathrm{WT}(t, f) = \int_{-\infty}^{+\infty} s(\tau)w_{k, p}(\tau - t, f; t)\exp{(-i2\pi f\tau)}\,d\tau\,, \label{eq:wt}
\end{equation}
where $w_{k, p}(\tau, f; t)$ is a Gaussian window function
\begin{equation}
    w_{k, p}(\tau, f; t) = \dfrac{1}{\sqrt{2\pi}\sigma_{k, p}(t, f)}\exp{\left(-\frac{\tau^2}{2\sigma_{k, p}^2(t, f)}\right)} \,, \label{eq:wt_windows}
\end{equation}
with a time-frequency-dependent standard deviation $\sigma_{k, p}(t, f)$ given by 
\begin{equation}
    \sigma_{k, p}(t, f) = k\left(f_0^p(t) + |f_0(t) - f|^p\right)^{1/p}\,,   
\end{equation}
where $f_0(t)$ is the dominant frequency of the signal and $k\in\mathbb{R}$ and $p\in\mathbb{R}$ are the scaling and spread factors, respectively. It is worth mentioning that when $p = 1.0$ in Eq. \eqref{eq:wt}, the canonical WT proposed by \cite{wtpaper} is recovered, while the change rate of the Gaussian
window with respect to the frequency can be regulated using $p>1.0$ as demonstrated by \cite{wang2022time}. In this regard, the free parameters in the WT (Eq.~(\ref{eq:wt})) adjust the shape of the Gaussian windows function (Eq.~(\ref{eq:wt_windows})) and can be used to optimize the amount of information present on the spectrograms~\citep{stankovic2001, tp_wt, fractwt, desouza2025}, such that in this work we adopt $k=2.0$ and $p=5.0$. Therefore, the WT differs from the conventional CWT by (i) employing a nonstationary window function, (ii) localising the spectral energy around the dominant frequency, and (iii) removing the singularity problem related to the zero frequency in the time-frequency spectrum. In particular, the CWT can be recovered from the WT by taking the limit at which the dominant frequency dependency disappears in Eq.~\eqref{eq:wt_windows} (up to a phase factor). 

The dominant frequency $f_0$ is the main ingredient of the WT, and its computation is a research problem of its own~\citep{dom_freq_1, dom_freq_2, barnes1992, poggiagliolmi2014, hu2022, yu2023}. Considering the instantaneous frequency $f_{\mathrm{inst}}$ directly as the dominant frequency $f_0$ may result in strong numerical noise that exceeds the Nyquist limit~\citep{wang2006}. To address this, we follow \cite{wang2022time} and estimate the dominant frequency as a weighted average of the instantaneous frequency of the signal, that is,
\begin{equation}
    f_0(t) = \frac{\rho\int_{-\infty}^\infty L(t - \tau)a_{\mathrm{inst}}(\tau)f_{\mathrm{inst}}(\tau)\,d\tau}{\int_{-\infty}^\infty L(t - \tau)a_{\mathrm{inst}}(\tau)\,d\tau}\,,
\end{equation}
where $\rho$ is a scaling factor and $L(t)$ is a Gaussian low-pass filter with spread parameter $\sigma = 10^{7}\times\Delta t$, with $\Delta t$ being the sampling rate in time. The instantaneous amplitude $a_{\mathrm{inst}}$ and $f_{\mathrm{inst}}$ are computed through the analytical representation of the signal using the Hilbert transform. Considering that the LCs we are interested in are primarily periodic signals, we adjust the scaling factor $\rho$ such that the mean dominant frequency over the observation period coincides with the mean frequency present in the frequency spectra. Finally, once we compute the CWT and WT maps, we normalize their coefficients by the maximum magnitude of each map, so that the strongest response in each map is set to unity. We use the resulting normalized maps throughout the text for discussion.

\section{Comparison between wavelet and W Transforms} \label{comparison}

\begin{figure*}[!htpb]
	\centering
	\includegraphics[width=3.5in]{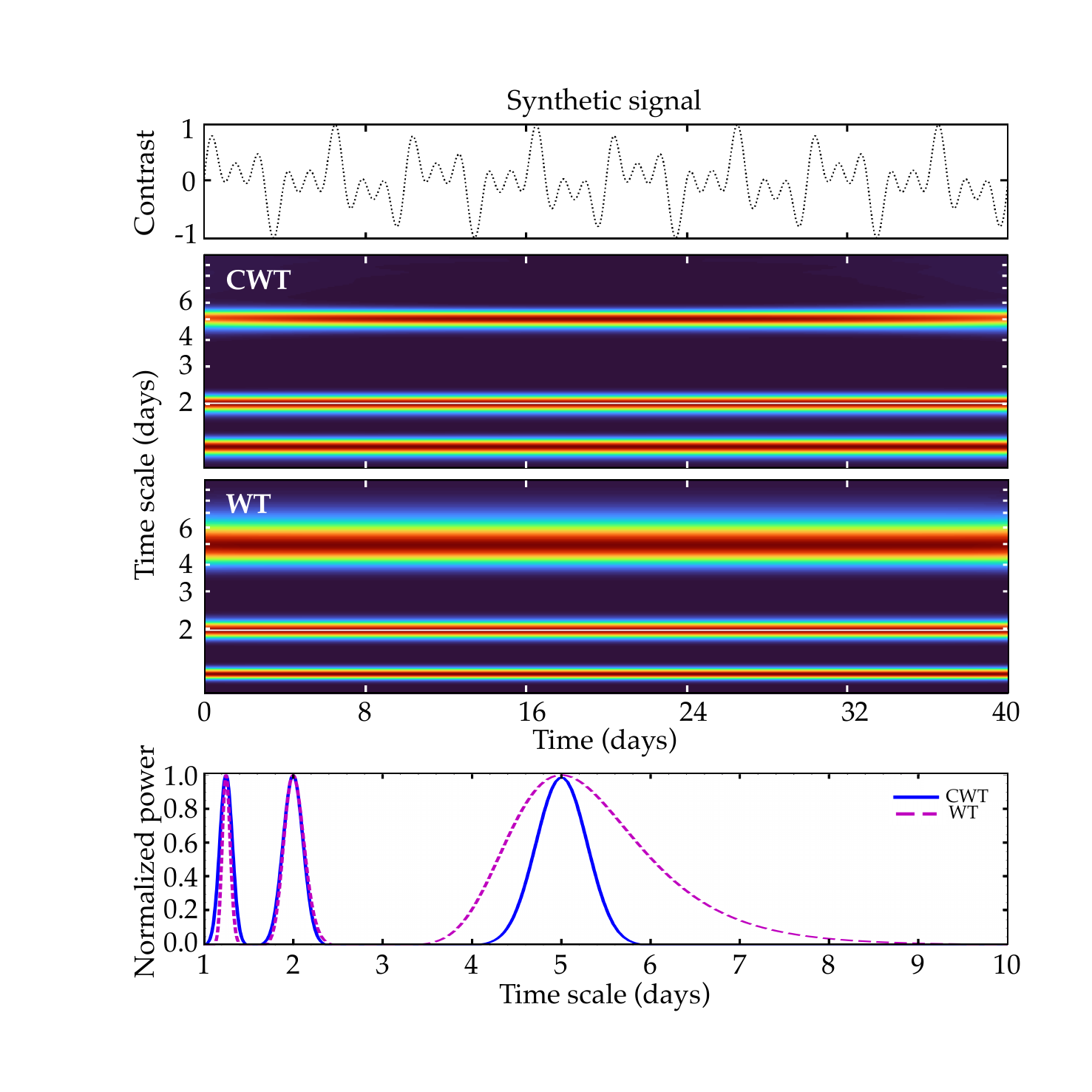}
	\includegraphics[width=3.5in]{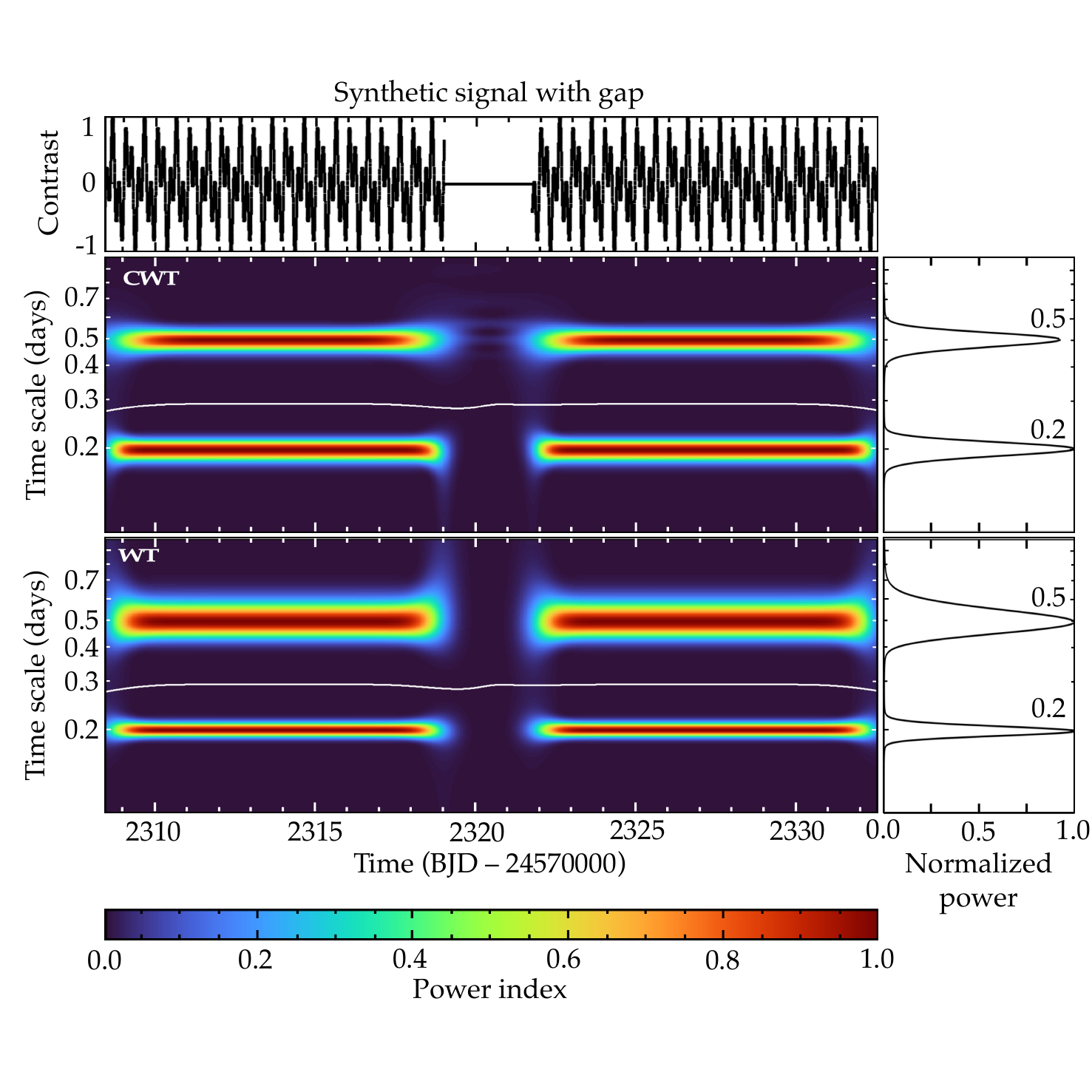}
    \caption{{\em Left panels}: Comparison between the WT and CWT for a synthetic periodic signal. The top three panels on the left show the time series, the CWT map, and the WT map, respectively. The bottom panel on the left presents the global spectra for the CWT (blue) and WT (purple). The global spectra are normalized by the maximum magnitude of each transform. The color palette in the maps represents the normalized power index levels, as indicated by the color bar at the bottom right. The solid white lines in the WT and CWT maps illustrate the dominant period, computed using the WT basis described in Section~\ref{transform}. {\em Right panels}: Comparison between the WT and CWT for a synthetic TESS-like LC with a typical gap. The top panel on the right shows the time series, while the panels in the second and third rows display the CWT and WT local maps and their respective global spectra. The numbers at each peak in the global spectra indicate their corresponding periods in days. The color palette and solid white lines in the maps follow the same conventions as those described for the left panels.}
	\label{fig:synthetics}
\end{figure*}

This study aims to apply WT for the analysis of periodic signals within typical timescales of stellar light curves. To illustrate the differences between WT and CWT in this context, Fig. \ref{fig:synthetics}, left panels, presents a pedagogical example featuring a noise-free synthetic signal composed of three superposed sine functions with the same amplitude of 1.0 and zero phase, but with periods of 1.25, 2.0, and 5.0 d. These periods were deliberately chosen because they are well separated in time scale and representative of the variability timescales of the stellar light curves analyzed in the following sections. The signal was generated using $5.0\times10^3$ time samples over 40 d, corresponding to a sampling interval of $0.008$ d. We stress that the Nyquist period is much shorter than the shortest period present in the signal. In this example, both the CWT and WT successfully identify the three period components of the signal. Their responses are nearly equivalent at shorter time scales, but the WT presents a sharp localization at the lowest period. For the longest-period component, both transforms correctly locate the dominant scale at approximately 5 days, although the WT distributes the associated power over a broader range of neighboring scales. This broader response reflects the WT's different time–scale localization and provides a more extended representation of the spectral content around the dominant scale.

In addition, this study applies WT to real-time series data, specifically to investigate stellar variability in TESS light curves. Although TESS data are mostly regularly sampled, they may contain long gaps that may hinder an accurate calculation of signal periods. The computational tools used in this work for both WT and CWT are optimized for regularly sampled time series, so the observational data must be interpolated to a regular sampling for proper analysis. To assess the effect of data gaps on the period recovery, we tested three simple gap-treatment approaches on simulated light curves with known periodic components: (i) linear interpolation, (ii) spline interpolation, and (iii) filling the gaps with a constant value corresponding to the quiescent level. For both CWT and WT, the interpolation methods tended to introduce artifacts in the time-frequency maps within the gaps and produced small shifts, typically of about 1--2\%, in the peaks of the corresponding global spectra. Indeed, simple interpolation methods are known to affect the recovered frequency content \citep[e.g.,][]{{Pascual-Granado2018}}. In contrast, filling the gaps with a constant value produced essentially no artificial signal within the gaps, while the peaks of the global spectra matched the injected periods without measurable shifts in our tests. More advanced methods, such as MIARMA \citep{Pascual-Granado2015}, could in principle be used to reconstruct the missing data while better preserving the frequency content. However, for the purpose of retaining signal information only where observations are available, the constant-filling approach was found to be sufficient. Figure~\ref{fig:synthetics}, right panels, shows an example of a synthetic signal composed of two sinusoids with equal amplitudes of 1.0 and zero-phase and periods of 0.2 and 0.5 d, generated with $1.0603\times10^4$ time samples over a time span of approximately 25 d (with a Nyquist period of about 0.0048 d), typical of a TESS sector, with a long gap in the middle. Similarly to the synthetic signal of left panel of Figure~\ref{fig:synthetics}, this signal has equal amplitude of 1.0, zero-phase, and is also noise-free. Filling the gap with a constant value enables both the WT and CWT to accurately recover the true periodicities of the synthetic signals. However, WT is particularly effective in preserving the signal amplitudes, as reflected by the normalized power levels in the global spectra. While WT maintains consistent power levels for both the 0.2 and 0.5 d signals, CWT shows a tendency for a reduced power level for the 0.5 d signal compared to its actual value. This tendency is caused by the fact that CWT is more sensitive to boundary effects than the WT~\citep{mallat, Arts2022}. 

To complement this comparative analysis, we generated a series of synthetic light curves, spanning a typical TESS sector, to investigate the combined effects of data gaps, noise, and simulated rotation period values on period recovery (see Supplementary Material). The first set (Figs.~\ref{fig:Gaps3A}–\ref{fig:Gaps12A}) was generated with fixed periods of 0.2 and 0.5 d, including examples with 3, 6, and 12 gaps and three Gaussian noise levels: high (0.05), intermediate (0.005) and low (0.00005), chosen to span a broad range of signal-to-noise conditions, including levels comparable to those observed in TESS light curves. The number of gaps was selected to mimic typical scenarios in TESS observations: early sectors often contained a single large central gap. In contrast, more recent ones commonly show three gaps, and the cases with 6 and 12 gaps reflect stars observed across two or four consecutive sectors. A second set (Figs.~\ref{fig:Gaps3B}–\ref{fig:Gaps12B}) of simulated light curves was constructed with the lowest noise level fixed, now considering 3, 6, and 12 gaps and period pairs ranging from 0.05-0.2 d, through 0.5-1.0 d, up to 1.0-5.0 d, covering representative values of stellar rotation periods. 

These additional sets of synthetic signals show some interesting aspects: in the first set (Figs.~\ref{fig:Gaps3A}–\ref{fig:Gaps12A}), the periodicities of 0.2 and 0.5 d could still be recovered with both WT and CWT, even in the presence of multiple gaps. However, the WT maps become progressively more fragmented as the number of gaps increases, while the CWT maps preserve clearer structures. The dominant effect arises from the noise level. In that case, WT maintains stable power levels for both periodicities across all cases, whereas CWT shows attenuation, particularly for the longer period (0.5 d) under high-noise conditions. We also note that, in these examples, the CWT maps show less power spread than the WT. This aspect indicates that, in our tests, gaps alone did not prevent period recovery. However, noise strongly amplified the differences between the two methods. In the second set (Figs.~\ref{fig:Gaps3B}–\ref{fig:Gaps12B}), the input periodicities could be recovered with WT and CWT for all combinations of periods, even in the presence of multiple gaps. In these cases, the 5.0 days peak is consistently weaker in CWT than in WT across the tested gap configurations. 

\section{The W Transform applied to TESS stellar light curves} \label{results}

This study applies the W-Transform method to a photometric time series of six astrophysical objects: a Brown Dwarf, an Ultra-Cool Dwarf, a Solar Twin, a $\gamma$ Dor pulsating star, a G-type star planet hosting, and an eclipsing binary, using data gathered from the TESS space mission. TESS, a space-based telescope, was launched in April 2018 and positioned in a highly elliptical orbit that completes an entire cycle every 13.7 days. The mission's primary objective is to provide uninterrupted photometric observations of a significant part of the celestial sphere, divided into 26 sectors measuring 24° $\times$ 96°. In monitoring designated targets, the satellite employs a 2-minute cadence (referred to as short cadence) and a 20-second cadence (known as fast cadence) throughout the approximately 27-day monitoring period in each sector. Additionally, TESS produces light curves with a 30-minute cadence from full-frame images (FFIs). To maintain consistency in our data analysis, we utilized the 2-minute short-cadence light curves with the quality flag set to 0, ensuring the use of the most accurate and reliable data. The TESS light curves are nearly evenly sampled, but may contain gaps typically related to data downlink, telescope operation errors, or cosmic-ray incidences \citep[e.g.,][]{colman2024}. Because the~WT and~CWT algorithms used in the present work only support evenly sampled data, the light curves analyzed in this section were re-binned to a constant cadence of 20~min. Short gaps ($\lesssim 1$~ d) were filled with linear interpolation, and long gaps ($\gtrsim 1$~ d) were filled with a constant value (corresponding to a quiescent flux level).

The following stars were selected for the present comparative analysis between the W- and the CW- Transforms: the solar twin TIC 276293046 \citep{lehmann2023}, the brown dwarf TIC 102076870 \citep{2015ApJS..219...33G}, the Ultra-Cool Dwarf TIC 359892714 \citep{fontinele2024}, the $\gamma$ Dor pulsator  TIC 164173105 \citep{gomes2024}, the eclipsing binary TIC 355151781 \citep{Prsa2022}, and the hosting planet star TOI 1136, that is TIC 142276270 \citep{Dai2023}. To highlight the details of the W Transform, we compare it with the CWT by presenting the results of the analysis in two scenarios: first, by analyzing stars with only one light curve, and then, stars with two light curves, but presenting differences in their structures. 

\begin{figure*}
	\centering
	\includegraphics[width=3.5in]{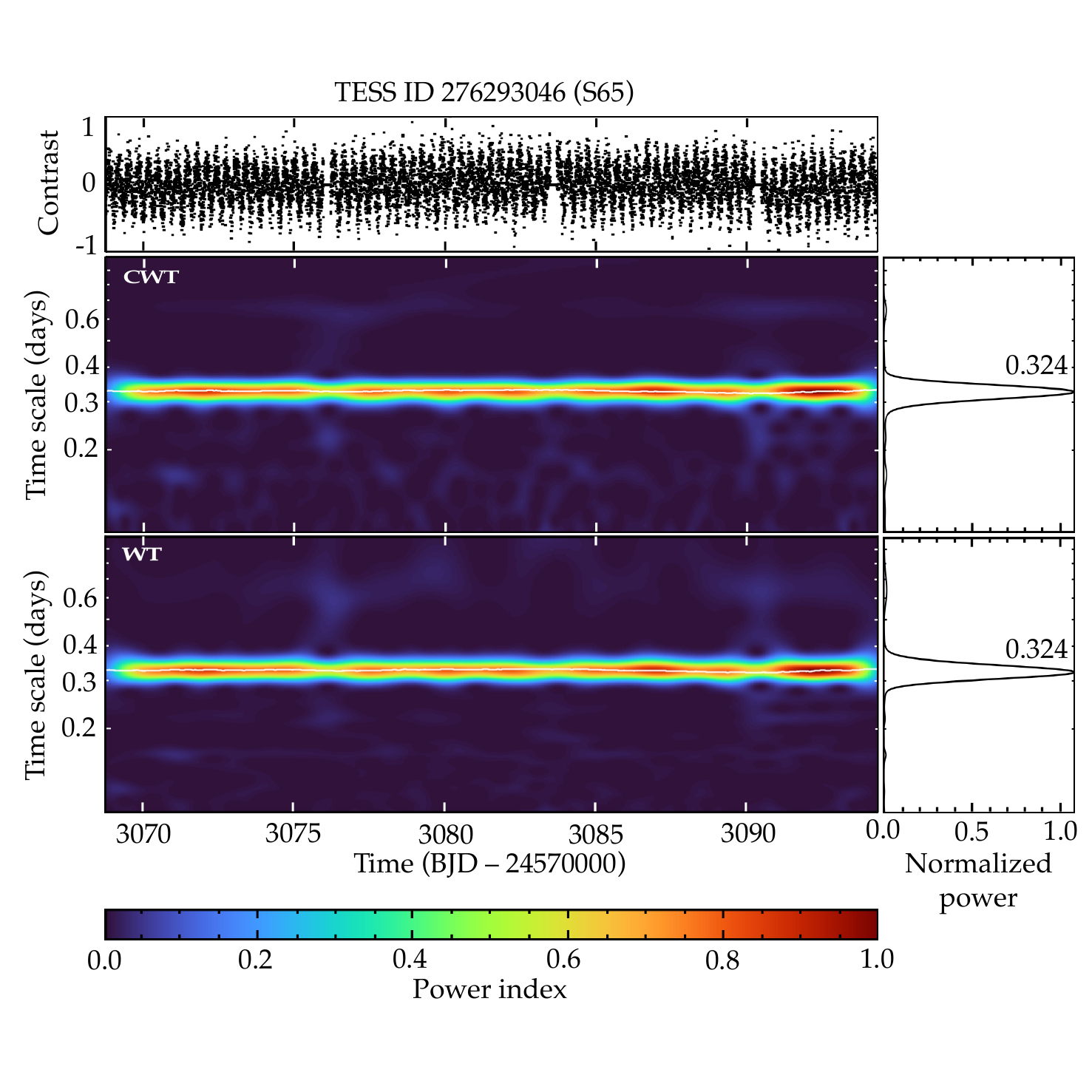}
	\includegraphics[width=3.5in]{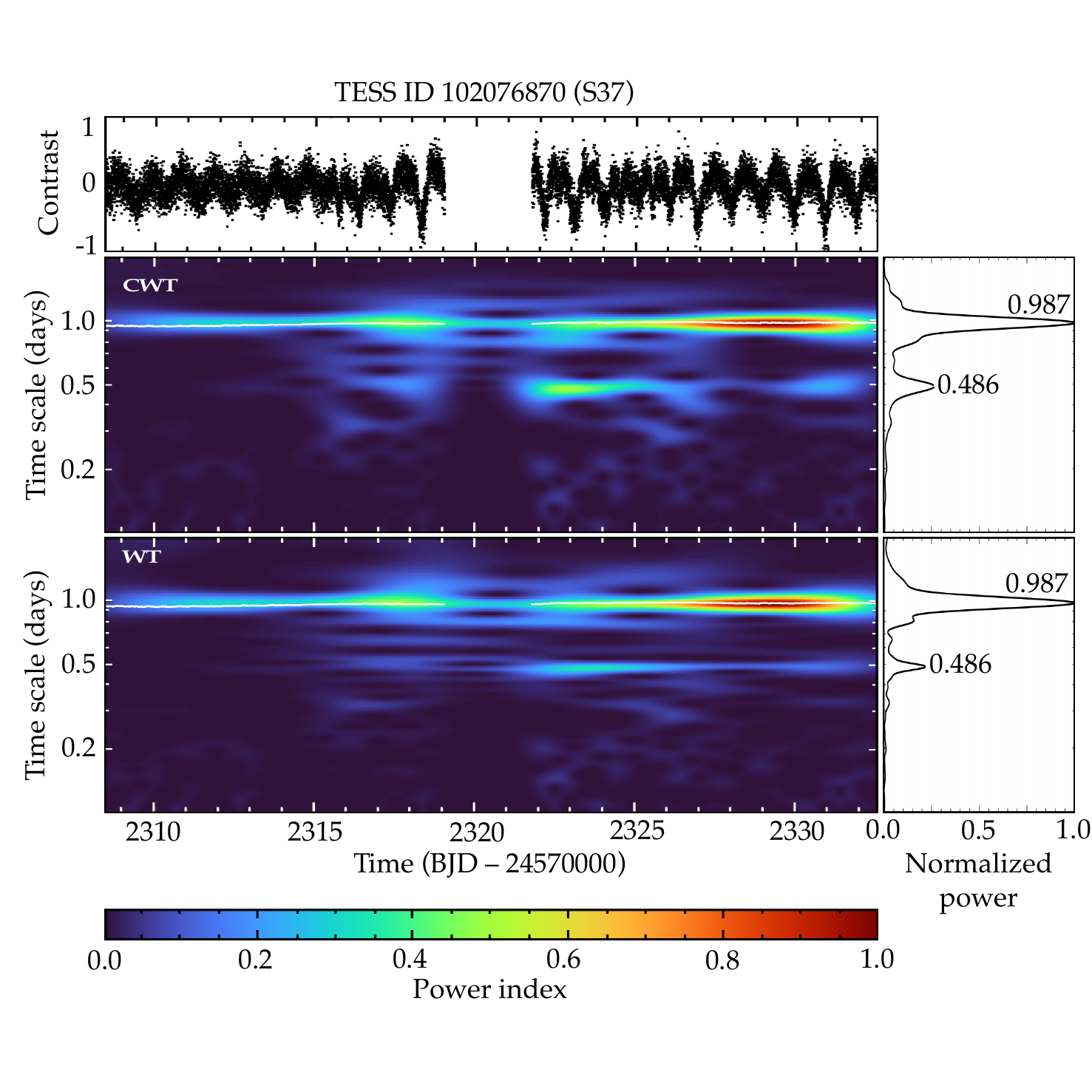}
    \includegraphics[width=3.5in]{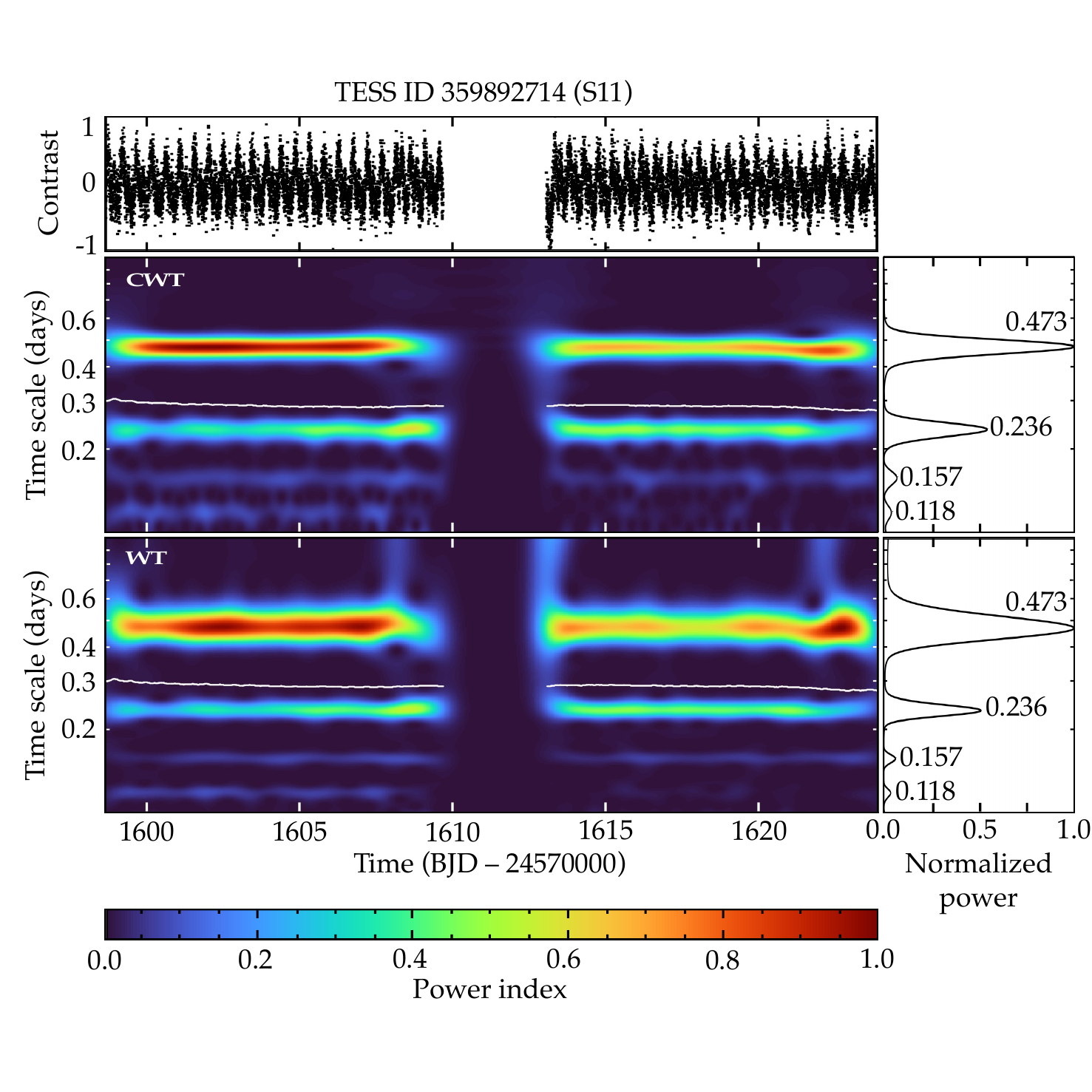}
	\includegraphics[width=3.5in]{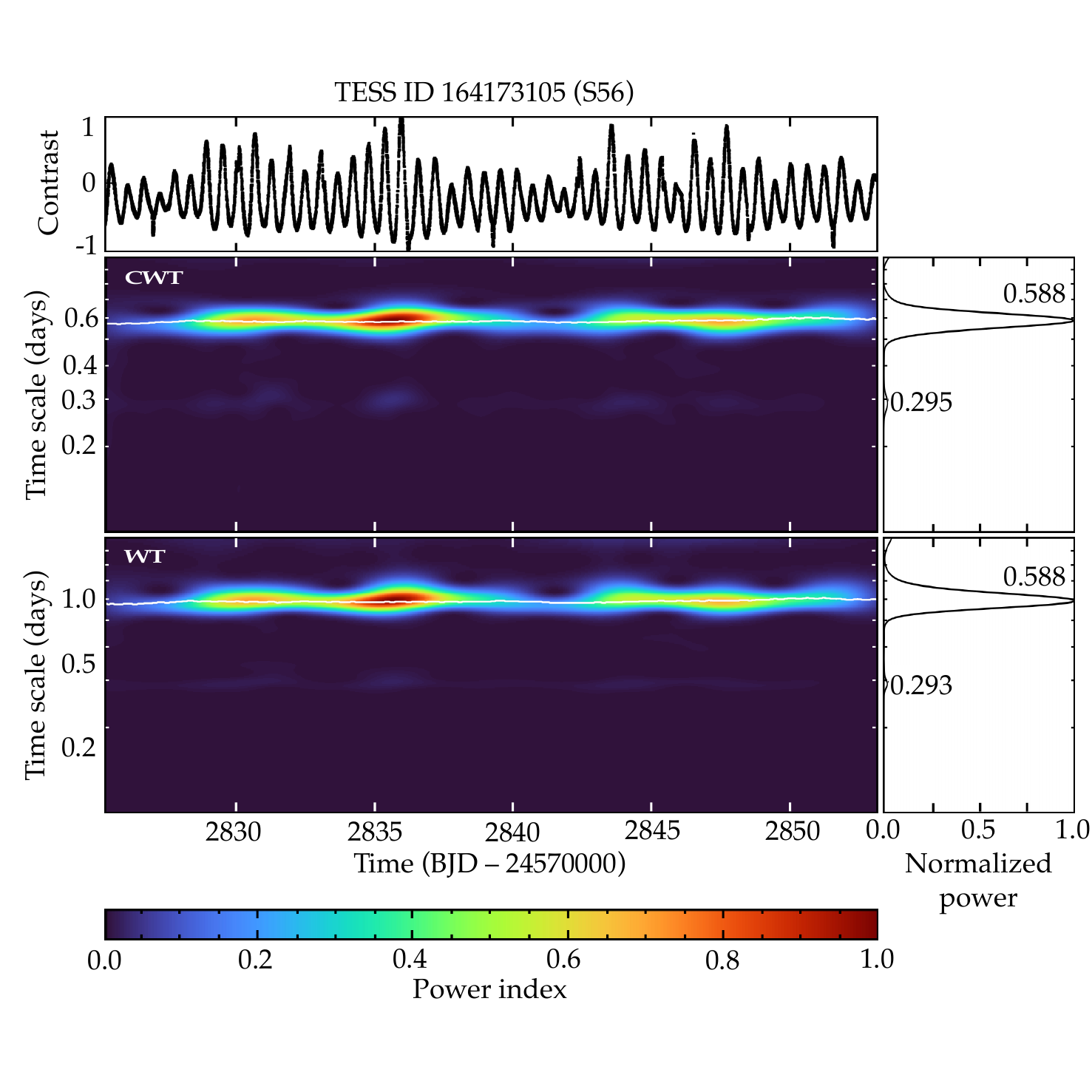}
    \caption{Comparison between WT and CWT methods applied to the TESS LCs. {\em Top-left panels}: Results for TIC~276293046, obtained from sector~65. {\em Top-right panels}: Results for TIC~102076870, obtained from sector~37. {\em Bottom-left panels}: Results for TIC~359892714, obtained from sector~11. {\em Bottom-right panels}: Results for TIC~164173105, obtained from sector~56. The layout of each set of panels, labeled with the TESS ID at the top, as well as the use of colors, lines, and symbols, follows the same conventions described for the panels on the right side of~Fig~\ref{fig:synthetics}.}
	\label{fig:results}
\end{figure*}

Figure~\ref{fig:results} displays the analysis of the WT and CWT for stars with only one light curve, namely,  the solar twin TIC 276293046, the brown dwarf TIC 102076870, the Ultra-Cool Dwarf TIC 359892714, and the $\gamma$ Dor pulsator  TIC 164173105. In each set of panels labeled with the TESS ID on top, the upper panel presents the LC, while the panels below the LC show the local map in the center and the global spectra on the right, with the CWT at the top and the WT at the bottom. The analysis of the stellar light curves in our sample using WT and CWT revealed significant findings.

\begin{figure}[ht!]
	\includegraphics[width=3.5in]{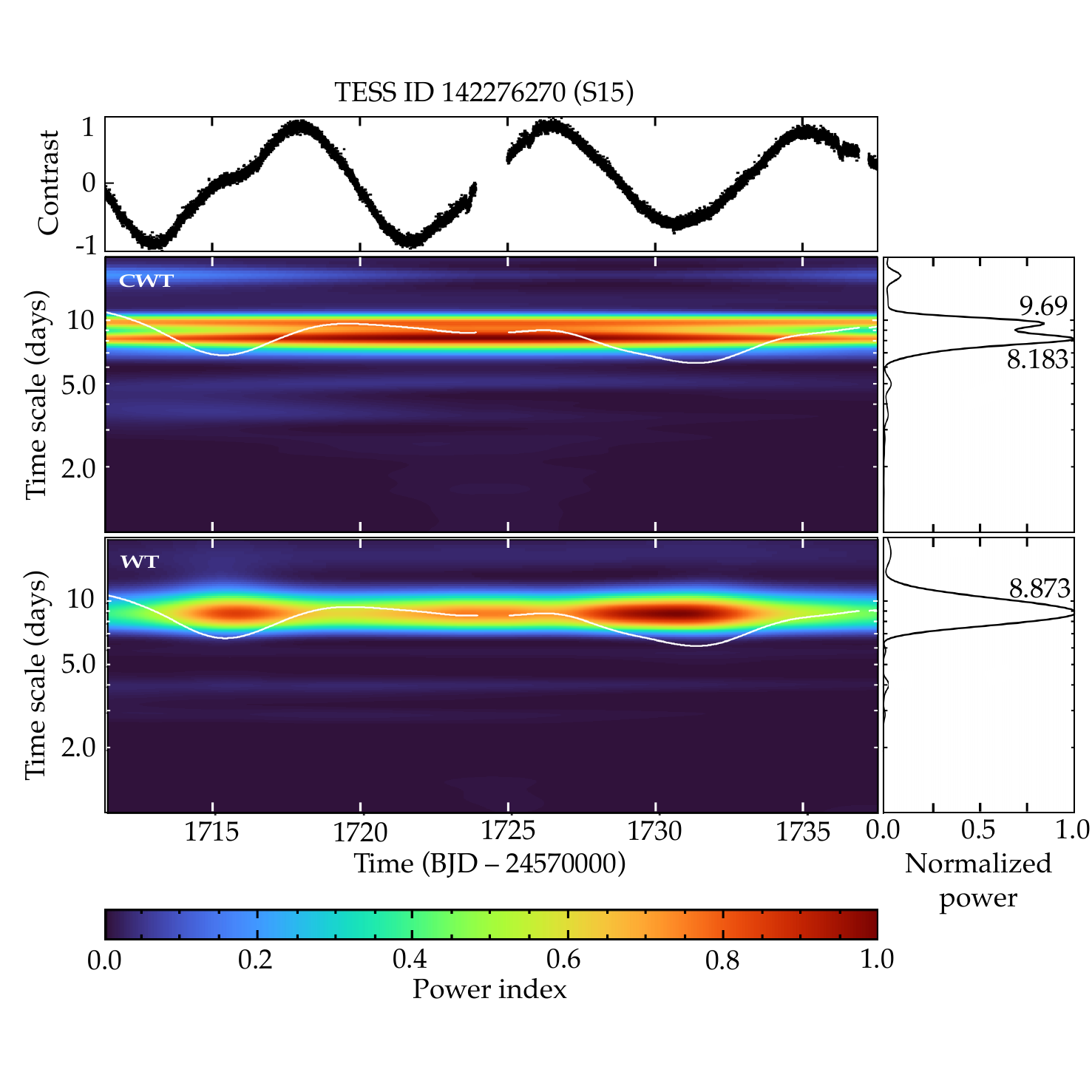}
	\includegraphics[width=3.5in]{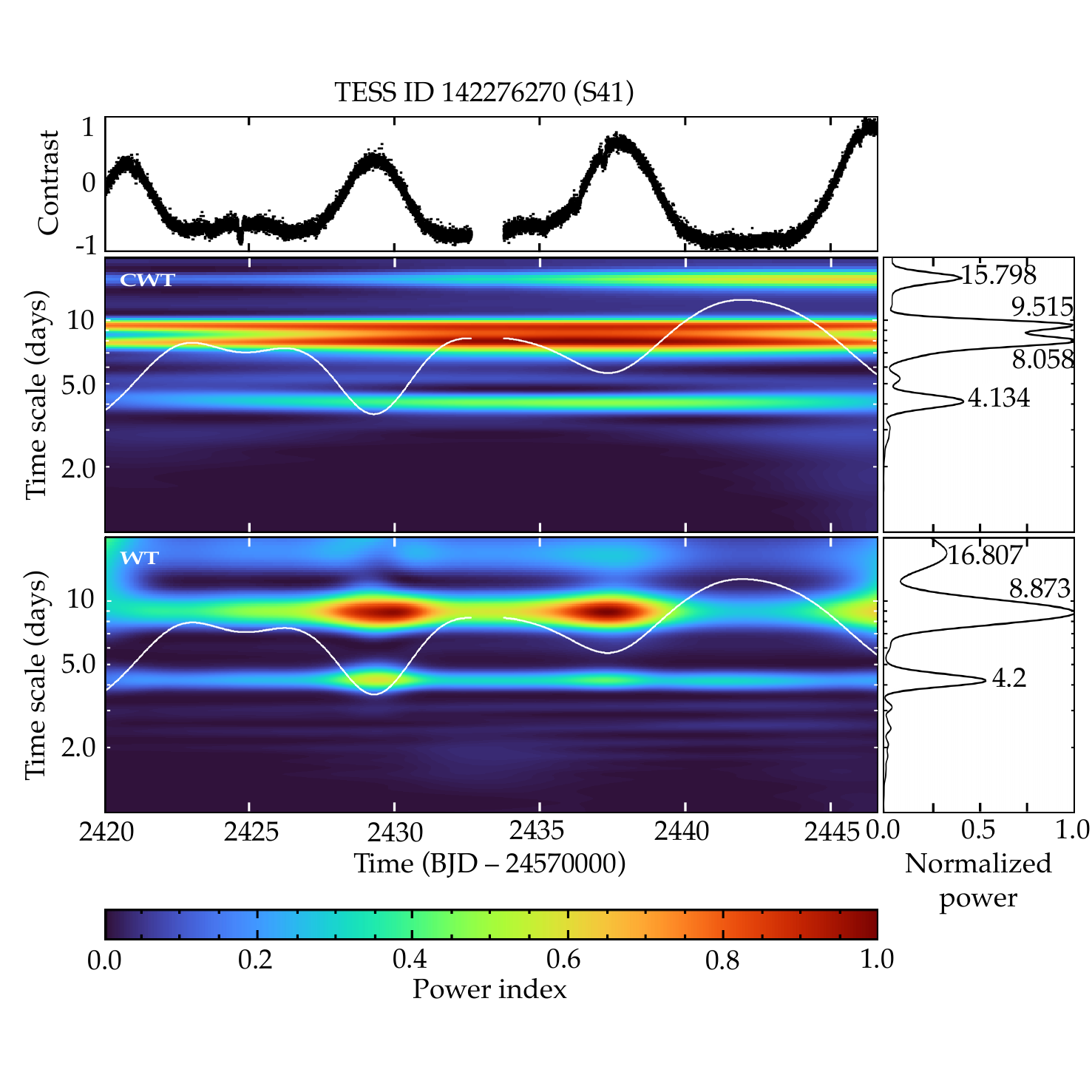}
	\includegraphics[width=3.5in]{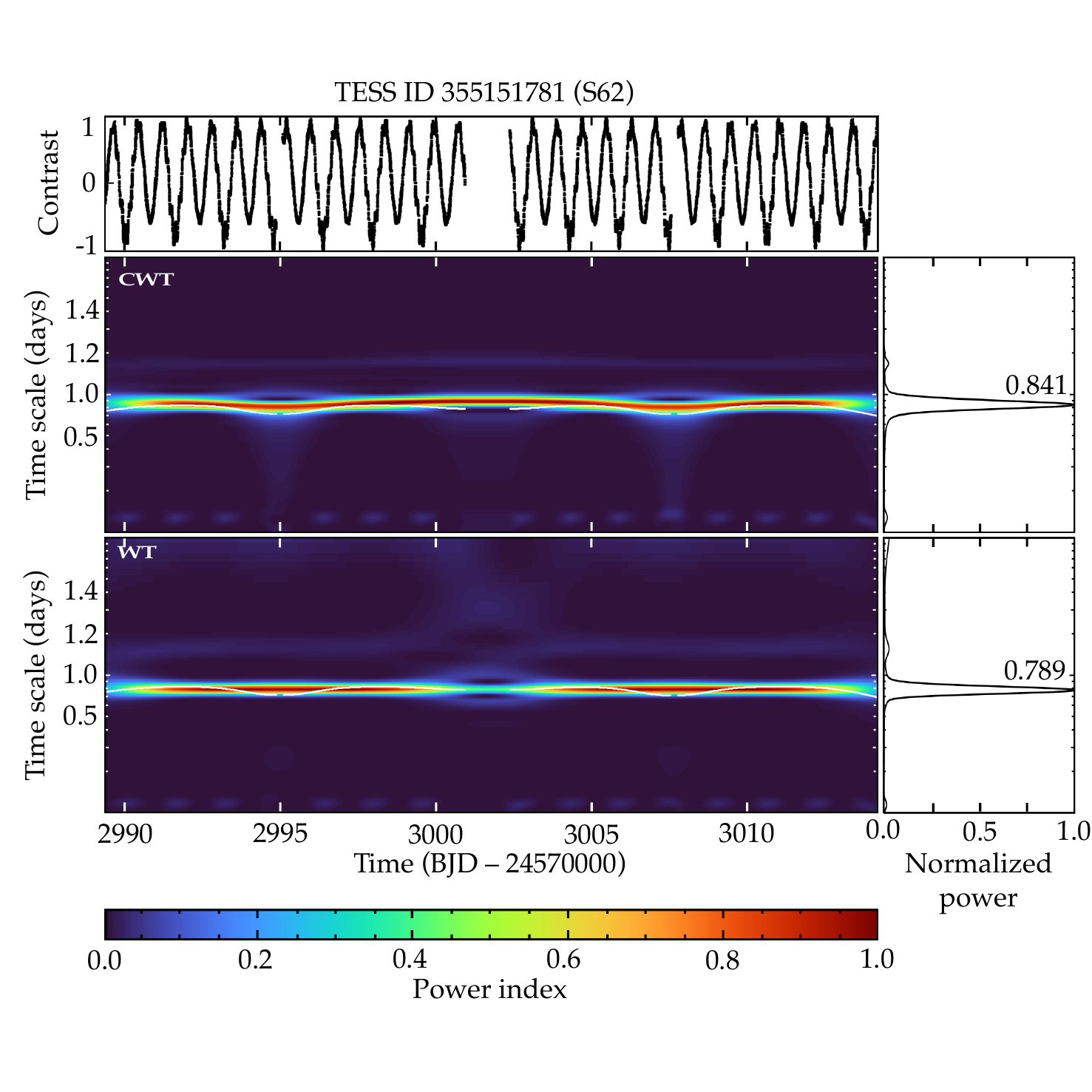}
	\includegraphics[width=3.5in]{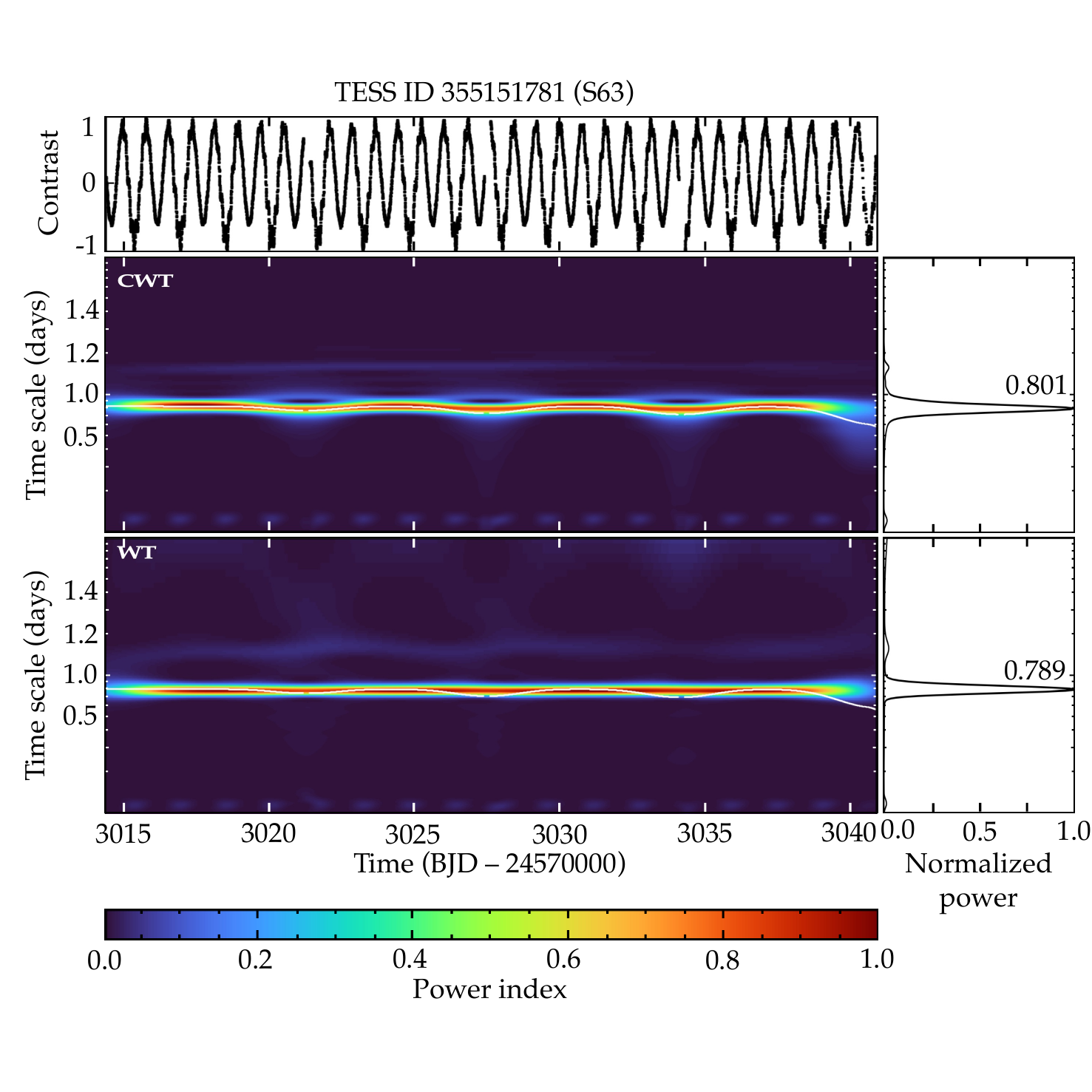}
\caption{Comparison between WT and CWT methods applied to TESS LCs. {\em Top panels}: Results for TIC~142276270, obtained from sector~15 (left) and sector~41 (right).{\em Bottom panels}: Results for TIC~355151781, obtained from sector~62 (left) and sector~63 (right). In each set, the top sub-panels show the light curves, while the middle and bottom sub-panels display the CWT and WT maps, respectively, together with their global spectra on the right-hand side. The layout of each set of panels, labeled with the TIC ID at the top, as well as the use of colors, lines, and symbols, follows the same conventions described for the panels on the right side of~Fig.~\ref{fig:synthetics}.\label{fig:TIC355151781_TIC142276270}}
\end{figure}

\begin{figure}[!htpb]
    \centering
    \includegraphics[width=3.5in]{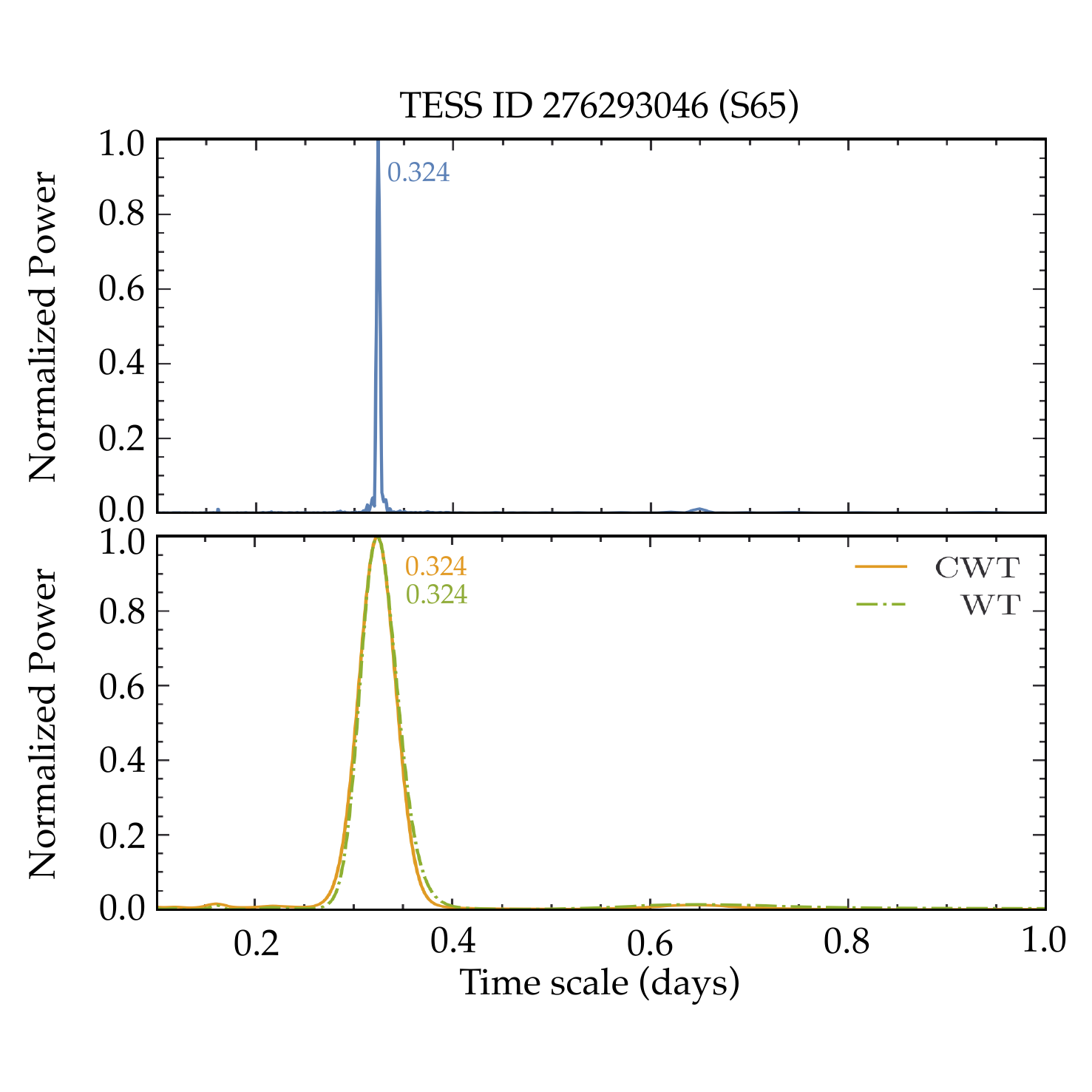}
    \includegraphics[width=3.5in]{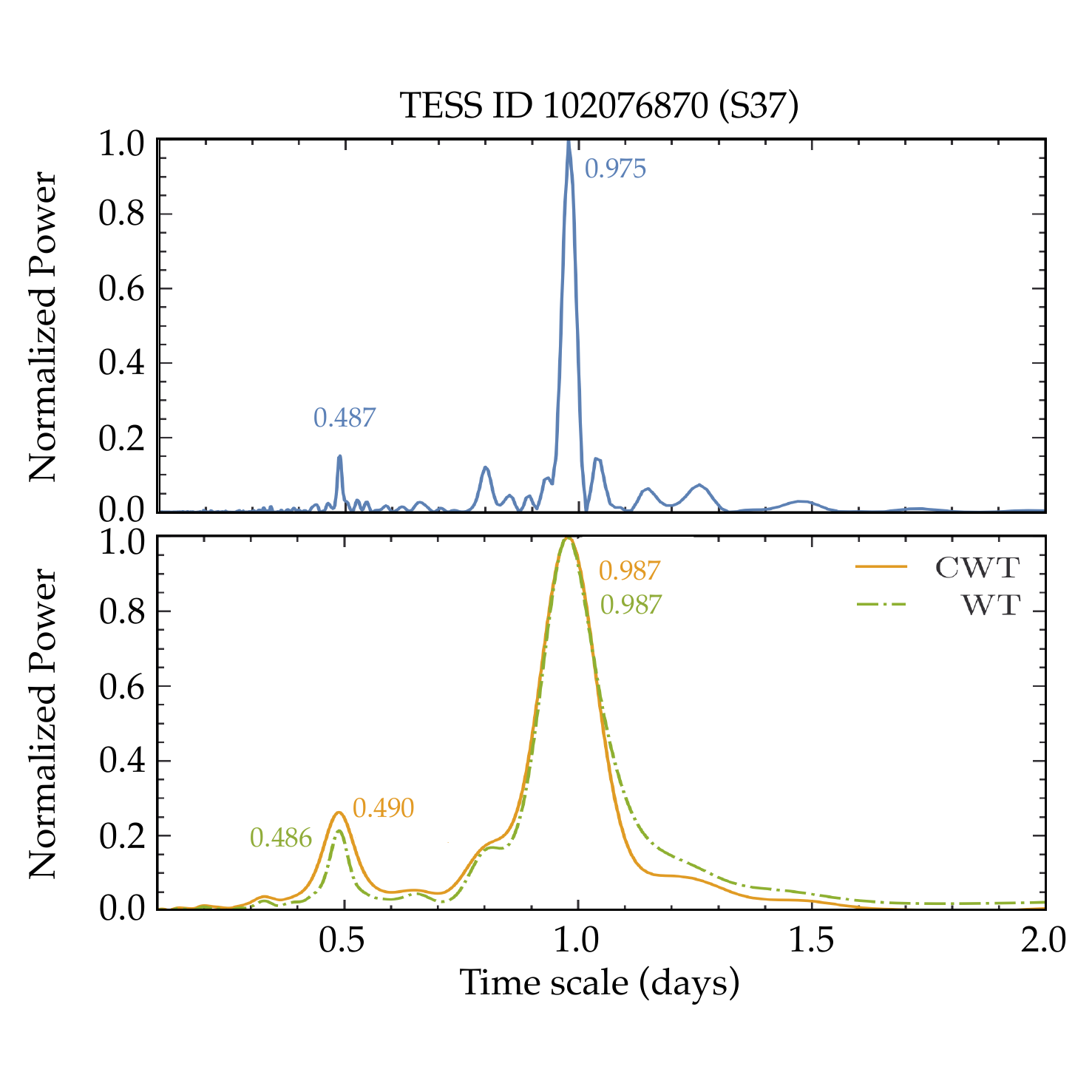}
    \includegraphics[width=3.5in]{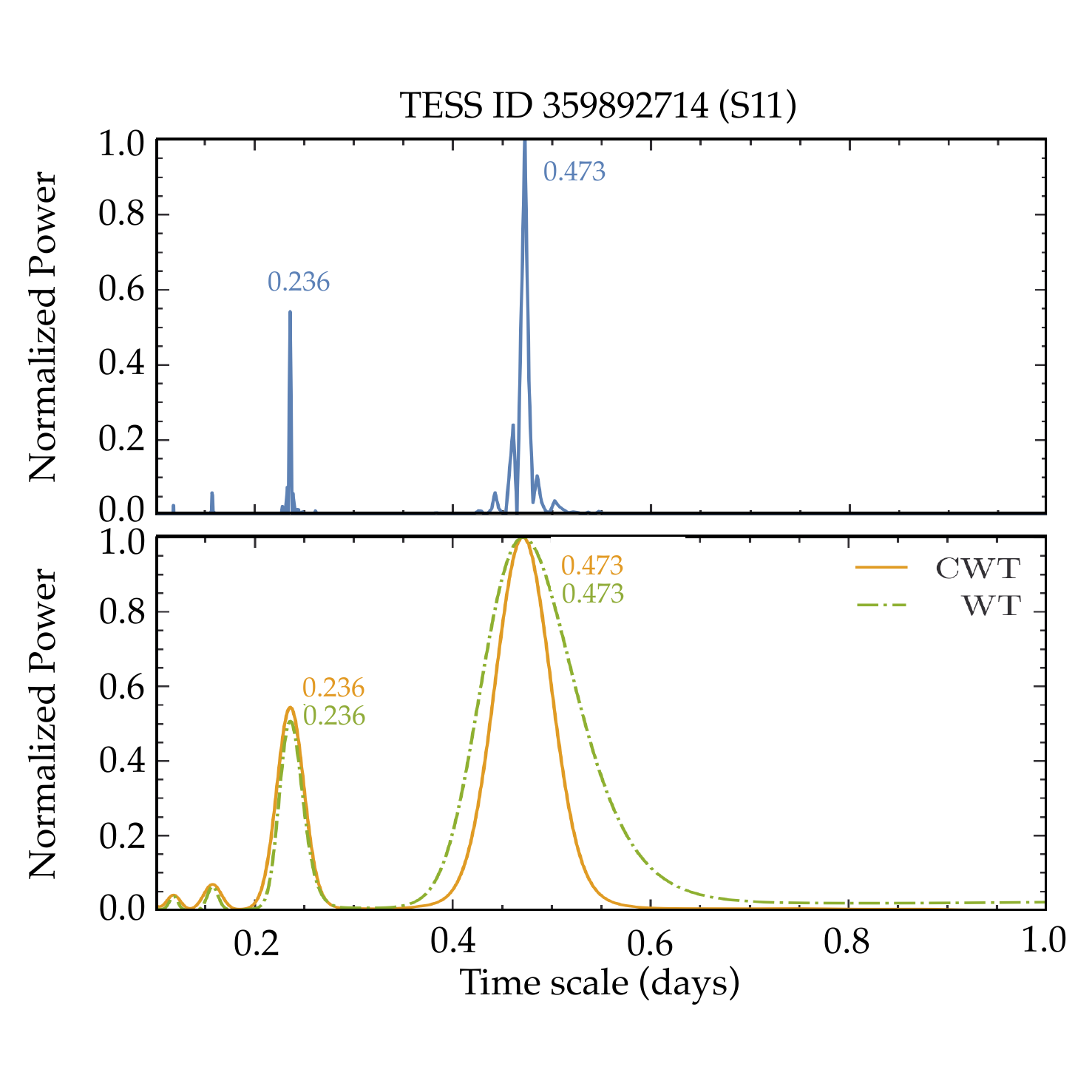}
    \includegraphics[width=3.5in]{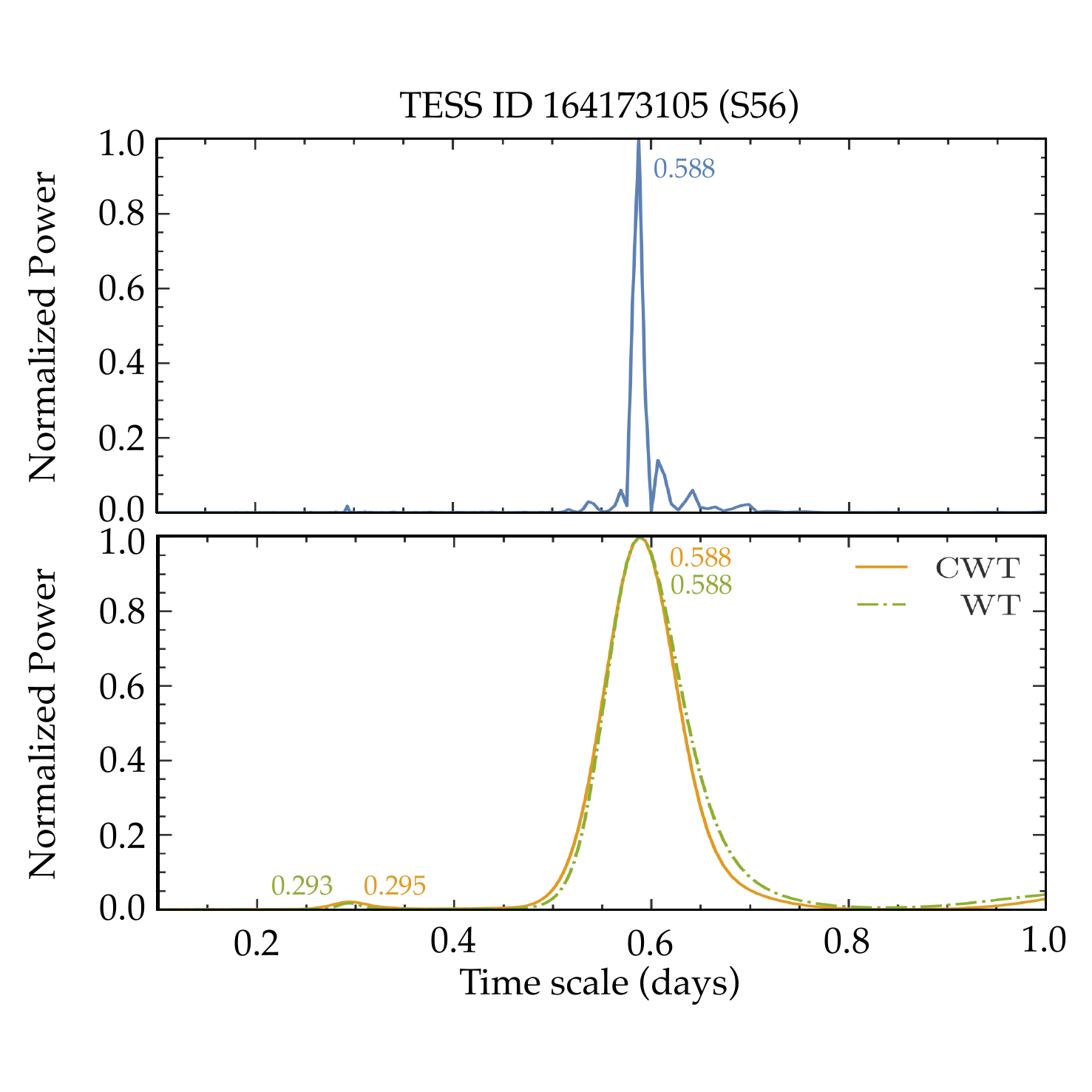}
    \caption{Comparison between the~WT and~CWT methods for the four TESS light curves analyzed in Fig.~\ref{fig:results}. Each set of panels, labeled with the corresponding TESS identifier, includes the Lomb-Scargle periodogram of the time series in the upper panel and the CWT overplotted with the WT global spectra in the lower panel. The line styles, as indicated in the legend, highlight the deviations between the~WT and CWT~methods. The numbers near the peaks indicate the corresponding peak periods in days.}
    \label{fig:ex2}
\end{figure}

\begin{figure}[!htpb]
    \centering
    \includegraphics[width=3.5in]{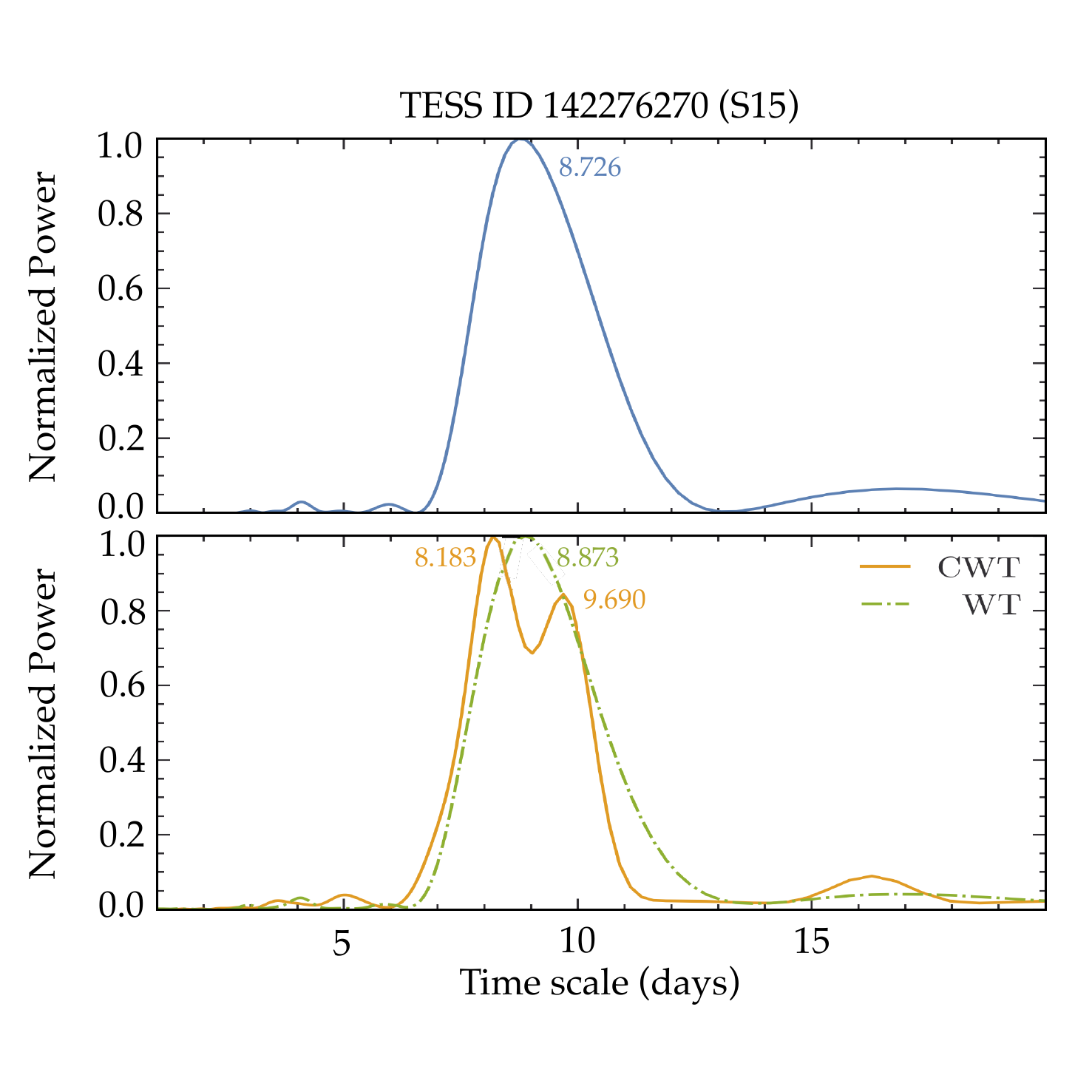}
    \includegraphics[width=3.5in]{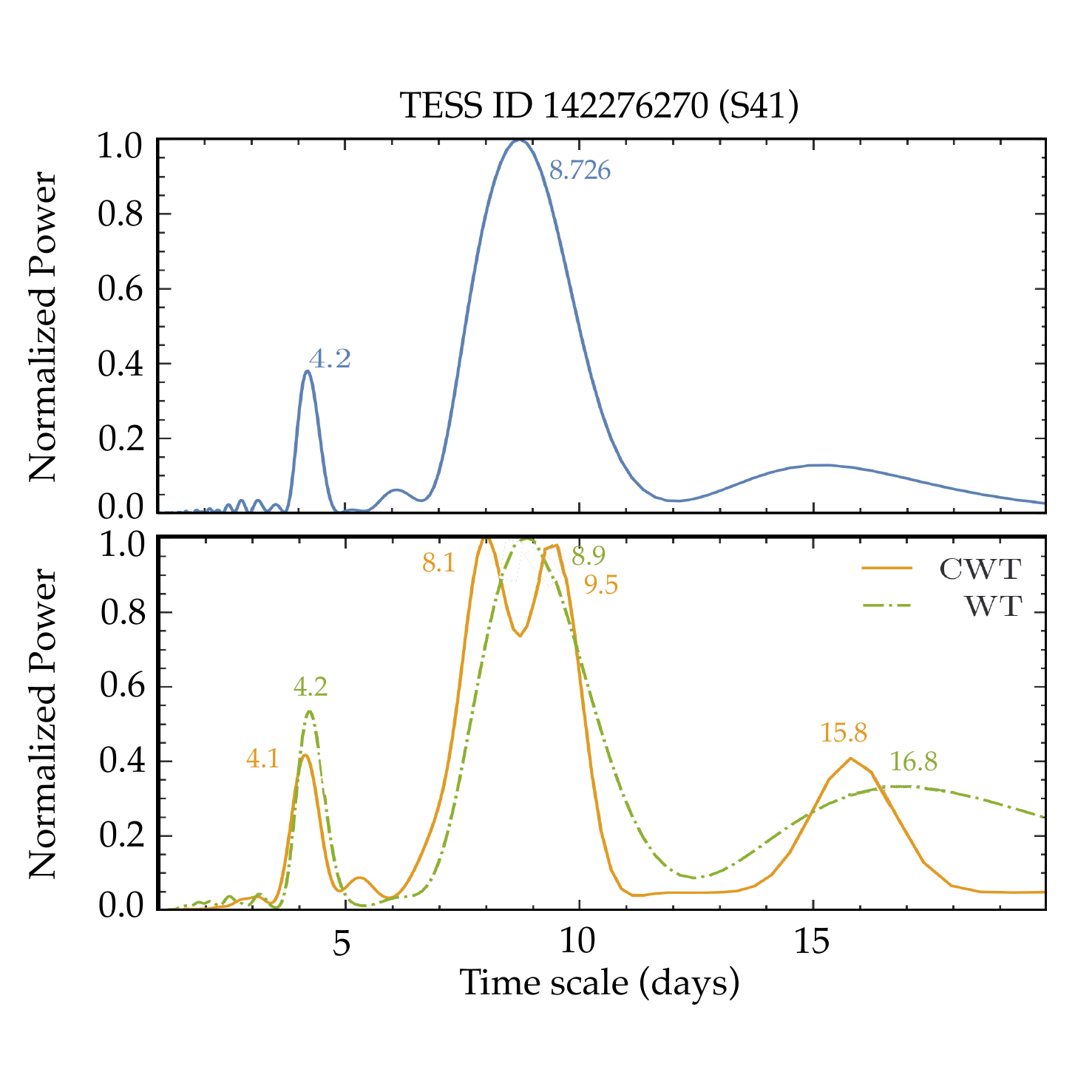}
    \includegraphics[width=3.5in]{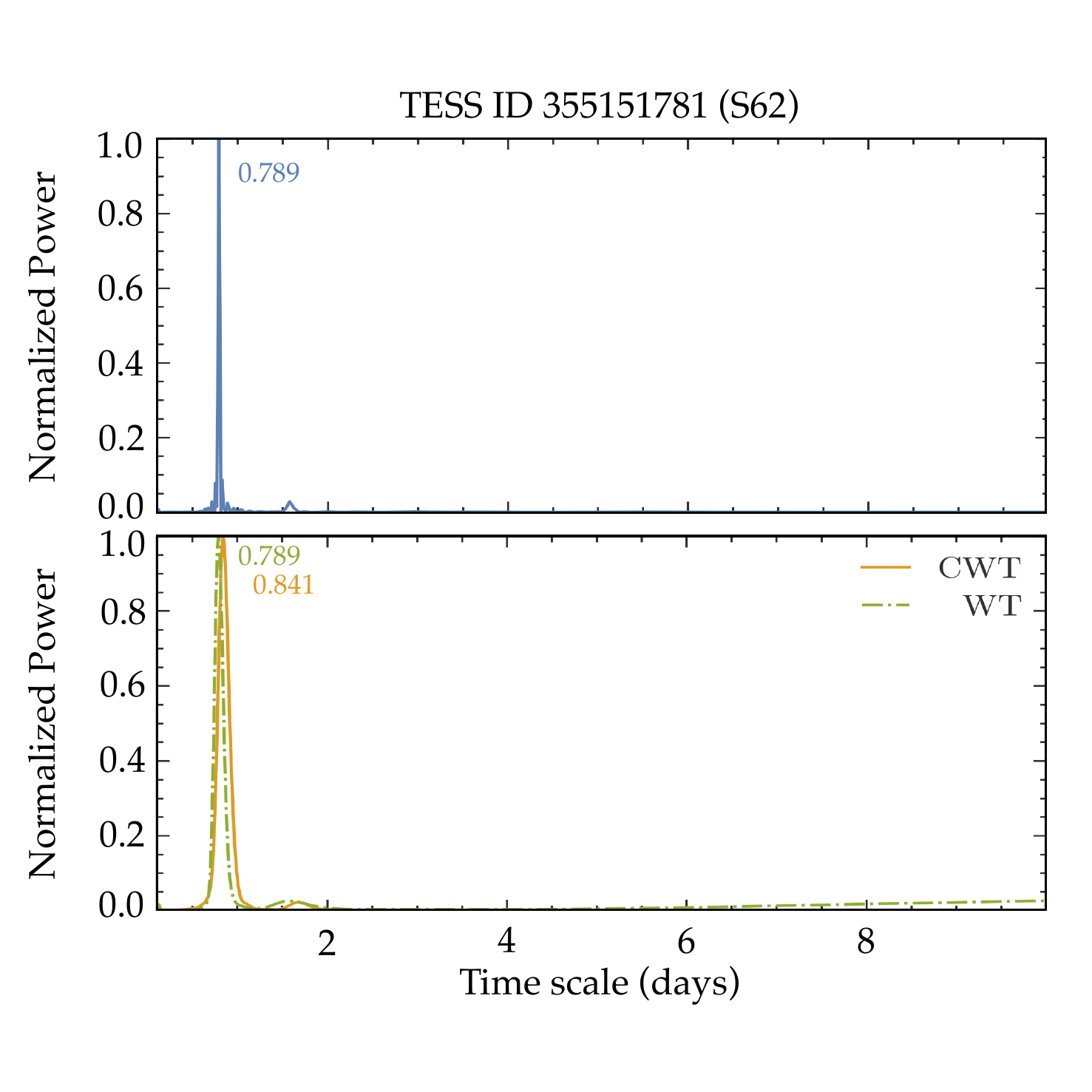}
    \includegraphics[width=3.5in]{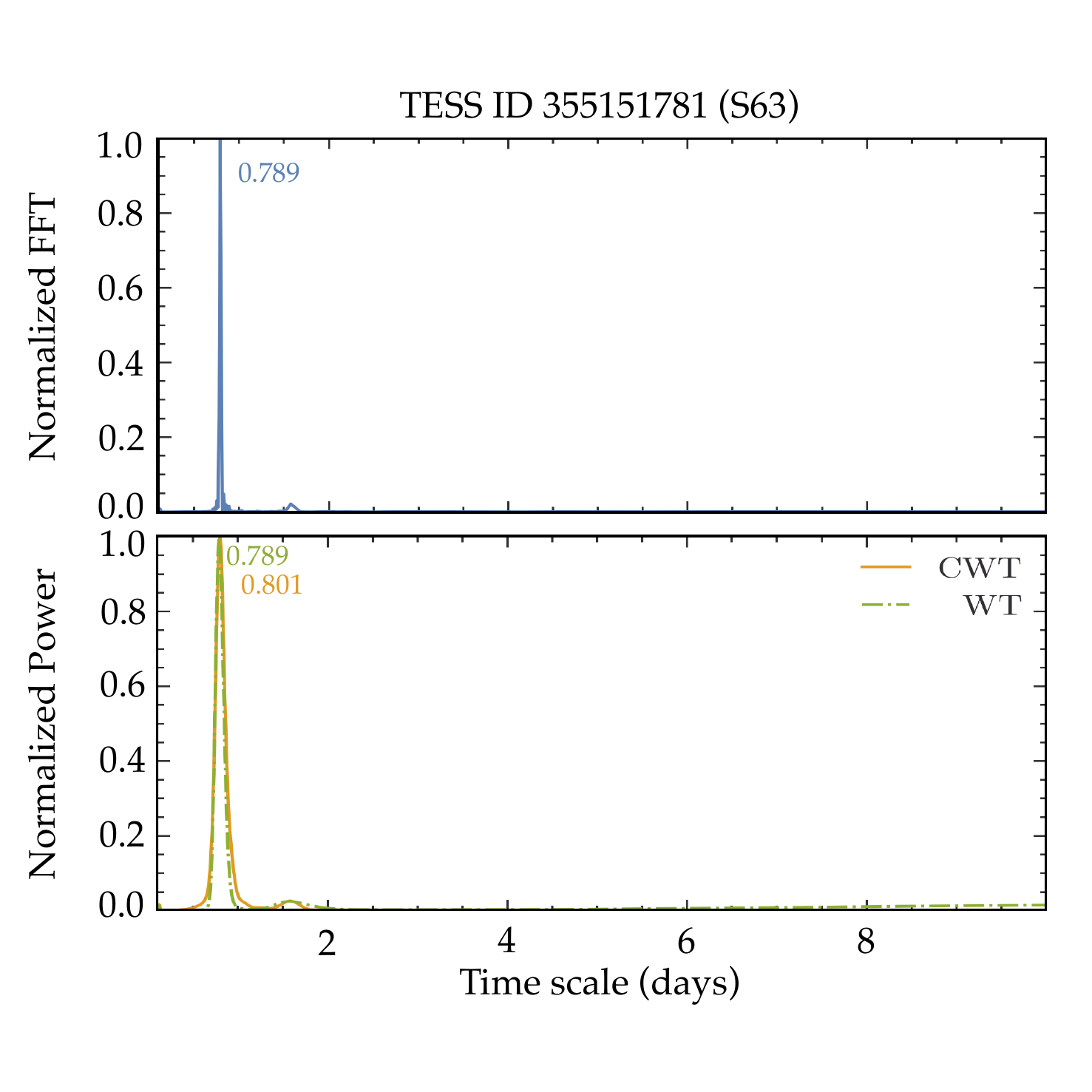}
    \caption{Comparison between the~WT and~CWT methods for the four TESS light curves analyzed in Fig.~\ref{fig:TIC355151781_TIC142276270}. Each set of panels, labeled with the corresponding TESS identifier, includes the Lomb-Scargle periodogram of the time series in the upper panel and the CWT overplotted with the WT global spectra in the lower panel. The line styles, as indicated in the legend, highlight the deviations between the~WT and CWT~methods. The numbers near the peaks indicate the corresponding peak periods in days.}
    \label{fig:global_2}
\end{figure}

The star TIC~276293046 is a solar analogue with an effective temperature $T_{\rm eff} = 6197 \pm 80$~K, a surface gravity $\log[g/({\rm cm~s}^{-1})] = 4.273 \pm 0.062$~dex, and a metallicity $[Fe/H] = 0.090 \pm 0.042$~dex \citep{lehmann2023}. The present work provides the variability period of this star for the first time, based on high-quality TESS light curves. Both WT and CWT global spectra of Fig.~\ref{fig:results}, top-left panels, give a main period of~0.324~days, as well as the Lomb-Scargle power spectrum, as shown in Fig.~\ref{fig:ex2}, top-left panels. The~WT and CWT~maps reveal the persistence of this signal throughout the entire time span. Overall, the signature displayed in the WT and CWT maps agrees with a rotational modulation pattern \citep[e.g.,][]{2020ApJS..250...20C} and the relatively fast rotation suggests that the star is a young Sun-like star \citep[e.g.,][]{leao2015}.

The brown dwarf TIC~102076870 hosts the first directly imaged planetary companion~\citep{chauvin2004} and has an estimated rotation period of about~1~day~\citep[e.g.,][]{scholz2005,stelzer2007}. This work presents a more precise determination of the rotation period of this star, based on TESS data. The global spectra shown in Fig.~\ref{fig:results}, top-right panels, reveal a main period of~0.987~days for both WT and CWT, which agrees with previous findings. These periods are consistent with the Lomb-Scargle period of 0.975 days, as displayed in Fig.~\ref{fig:ex2} (top-right panels), with a deviation of 2\%. There is also a half-period second peak in the~WT and~CWT global spectra that is typically associated with an asymmetric spot distribution in the stellar surface that may produce double-dip segments per rotation \citep{basri2018}. In the considered time series, the second-peak signature occurs within a specific time interval of the~WT and~CWT maps, approximately from BJD~24570002315 to the end of the light curve. For this time series, the features present in the~WT~map look sharper than the corresponding features present in the~CWT~map, especially at high frequencies (short periods).

The Ultra-Cool Dwarf TIC~359892714, recently analyzed by \citet{fontinele2024}, presents in the WT and CWT global spectra of Fig.~\ref{fig:results} (bottom-left panels), the main period of 0.473~days, associated with the star's rotation. This value is consistent with the corresponding Lomb-Scargle period of 0.446~days (see Fig.~\ref{fig:ex2}, bottom-left panels), with a deviation of about 6\%. The~WT and~CWT~maps reveal a double-dip signature \citep{basri2018}, with the main and half periods showing persistence throughout the entire time range and appear as two firm peaks in each global spectrum. An important nuance is that the high-frequency (short-period) features are better resolved in the WT map, while low-frequency (long-period) features are more contrasted in the CWT map.

TIC~164173105 was recently studied from~TESS light curves by~\citet{gomes2024} and classified as a $\gamma$~Dor pulsator with a pulsation period of~0.57~days. Both~WT and~CWT global spectra of~Fig.~\ref{fig:results} (bottom-right panels) exhibit a main period of~0.587~days, which is persistent in the~WT and~CWT~maps throughout the entire time span. This period agrees with the corresponding Lomb-Scargle period of~0.585~days (see Fig.~\ref{fig:ex2}, bottom-right panels) within a deviation of less than~1\%.

\subsection{Stars with multiple gaps and different structures in the LCs}{} \label{sec:tessdifferentstructures}

In the previous examples, we have considered only stars observed in a single sector, from where it is not possible to identify the effects of time-dynamic changes in the LC of the same star, namely, a star observed in multiple sectors. The following stars, TIC 142276270 and TIC 355151781, which benefit from multiple available TESS sectors, allow us to analyze these effects, including, in particular, changes in the LC structure and in the number of gaps with time.   

TIC 142276270, namely TOI 1136, is a G-type star that hosts at least six transiting planets with radii between $\sim$2 and 5 R$_{\oplus}$ \citep{Dai2023}. For this target, we analyzed LCs from two different TESS sectors (15 and 41, Fig.~\ref{fig:TIC355151781_TIC142276270}, top panels). The LCs show quasi-periodic variability patterns, but with a marked difference in their structures. In Sector 15 (top-left), the LC shows well-defined peaks and valleys. In contrast, Sector 41 exhibits well-defined peaks and wider valleys, introducing, in principle, more complexity to its structure. The global spectra of WT and CWT (Fig.~\ref{fig:global_2}, upper row) clearly reflect the effects of these differences. In both sectors, the WT retrieves a single main period of 8.873~d, persistent throughout the entire time span and consistent with the Lomb-Scargle result. This value is also in agreement with the independent determination by \citet{Dai2023}, who measured a rotation period of $8.7\pm0.1$~d. Although the CWT global spectra display two nearby peaks (8.183 and 9.690~d in Sector 15, and 8.058 and 9.515~d in Sector 41), this introduces ambiguity in determining the periodicities. While the CWT maps present a more continuous and coherent power distribution, the WT maps appear more fragmented, particularly in Sector 41, where the presence of widened valleys manifests itself as additional complexity in the maps. This fragmentation, together with the single main period recovered by the WT global spectra and the Lomb-Scargle periodogram, is a good indication that there are two close, unresolved peaks, which is precisely what is seen in the CWT maps. This behavior is consistent with the broader picture suggested by the synthetic tests, where the WT tends to preserve a unique and stable periodicity even under variations in the structure of the LCs. At the same time, the CWT shows greater sensitivity to these variations. This example also illustrates that the WT is particularly well suited to track time variations of the dominant frequency, which is where the method really excels, as shown by the structures in the upper right panel of Fig.~\ref{fig:TIC355151781_TIC142276270}.

TIC 355151781 is an eclipsing binary with an orbital period of 1.5807272~d \citep{Prsa2022}. For this target, we have analyzed LCs from two different TESS Sectors (62 and 63, Fig.~\ref{fig:TIC355151781_TIC142276270}, bottom panels), both LCs showing deep and narrow eclipses, recurring regularly, but with differences in the number of events effectively sampled (missed or partial eclipses). Indeed, the increase in the number of gaps directly reduces the number of detected eclipses, i.e., the event frequency, an aspect that can change the time-frequency content of the LC. As observed from the global spectra, both WT and CWT recover the main period associated with the orbital signal (visible at the strong harmonic $P/2 \approx 0.79$~d). However, the WT retrieves the same period of $\sim0.789$~d in both sectors. At the same time, the CWT peaks differ slightly, from 0.841~d in Sector 62 to 0.801~d in Sector 63, illustrating that the CWT is more sensitive to differences in the gap distribution and LC structure. This result is also consistent with synthetic tests, which showed that the WT preserves amplitudes and remains stable with the same periodicities, even when the effective number of events decreases. In contrast, the CWT is more vulnerable to distortions introduced by changes in the number of gaps.

\section{Conclusions} \label{sec:conclusion}

This work presents a time-frequency study of TESS LCs using a relatively novel method, the W Transform, which here is explored for the first time in the treatment of astrophysical data, which is proposed as an improvement over the Continuous Wavelet Transform. Similarities and differences between these two methods are explored, particularly in the context of periodic signals, using both synthetic and observational data. A comparison with the Lomb-Scargle method was also performed for reference. The observational data consisted of TESS LCs of six stars, namely, the solar analog TIC~276293046, the brown dwarf TIC~102076870, the ultra-cool dwarf TIC~359892714, the $\gamma$~Dor pulsator TIC~164173105, the multi-planet host TIC~142276270, and the eclipsing binary TIC~355151781.

Overall, the global and local spectra computed from both methods are similar in most cases. However, significant differences were identified for stars with LCs presenting sector-to-sector changes in structures or in the number of gaps, thereby revealing time-dynamic effects of the stellar variability, which, in particular, may influence the frequency of events associated with the corresponding LCs. These latter aspects are revealed by synthetic LCs and confirmed by observational data, with WT spectra being more stable and more accurate in recovering signal amplitudes. In contrast, for stars with LCs that exhibit different structures or changes in the number of gaps, the CWT maps show some distortions, particularly at the low- and high-frequency extremities, and also present limitations in recovering signal amplitudes. In essence, for a moderate noise level, CWT works better than WT with more gaps, and for a high noise level, WT works better than CWT with less gaps. Indeed, for observational data, the WT method resolved high-frequency (short-period) signal features with higher details, while the CWT method performed better in resolving low-frequency (long-period) features. This aspect indicates that both methods may complement each other in an in-depth analysis of a periodic astrophysical signal. Therefore, the inherent flexibility of the WT method in adjusting the dominant frequency could enhance the time resolution even within LCs where the CWT method performs better. The dominant period does not always coincide with peaks in the global spectrum. This is because the dominant period is a local quantity, corresponding to the scale of maximum wavelet power at each instant. In contrast, peaks in the global spectrum locate the scales of maximum time-averaged power. These two quantities coincide only when the signal is stationary and has a single periodicity; in general, they differ because the power distribution is time-averaged and because the wavelet power spectrum is known to be biased toward longer scales \citep{Liu2007}. In this context, our results suggest that the WT is also a robust and accurate tool for analyzing stellar LCs, offering refinements over the CWT and expanding the range of numerical techniques available for analyzing different stellar periodic phenomena from photometric observations. 

As previously highlighted, the spectacular increase in the number of stars with observed photometric light curves provided by the TESS space mission is greatly expanding our knowledge of stellar dynamics. However, this plethora of information requires a tremendous effort in data processing, including the identification and implementation of new computational procedures. At the same time, such richness of data, covering tens of millions of light curves, has revealed a great complexity still unknown, including the temporal behavior and persistence of associated phenomena, as well as unexpected effects related to instrumental issues such as gaps and noise levels. Given this situation, and inspired by the present study, a comparative study between different wavelet transforms reported in the literature, including the weighted wavelet Z-transform \citep{Foster1996} and the synchrosqueezed wavelet transform \citep{Daubechies2011, Ramon-Ballesta2022}, may open new horizons for the treatment and analysis of photometric light curves. In particular, such a study could help identify the persistence level of phenomena associated with specific variabilities and determine whether the detected frequencies and amplitudes are stable. Although both transforms are based on the classical wavelet formalism, they incorporate improvements that can make them complementary in the treatment of the referred complexities. For instance, the weighted wavelet Z-transform considers the wavelet not just as a projection onto a set of trial functions, but as a weighted projection; such a strategy increases its ability to detect frequency and amplitude fluctuations and to estimate their parameters with improved precision and consistency. The synchrosqueezed wavelet transform was proposed to reduce the width of the frequency bands of the classical wavelet transform by reallocating the energy around a central frequency while preserving time resolution, and providing a better resolution in frequency. Given the complexity revealed by TESS data, robust time-frequency analysis tools, such as those discussed here, will be essential to explore the forthcoming PLATO space mission \citep{Rauer2025}, which is expected to represent a new gold mine of photometric data, also producing tens of millions of light curves for different astrophysical objects.

\begin{acknowledgments}

Research activities of the Board of Observational and Instrumental Astronomy at the Federal University of Rio Grande do Norte (NAOS) are supported by continuous grants from the Brazilian funding agency CNPq.
This study was financed in part by the Coordenação de Aperfeiçoamento de Pessoal de Nível Superior -- Brasil (CAPES) -- Finance Code 001, and by the program CAPES/Print.
P.D.S.L. acknowledges a UFRN/CAPES/Print fellowship, process 88887.838338/2023-00.
D.O.F. acknowledges a CAPES graduate fellowship, and R.L.G. acknowledges a CNPq PDE fellowship.
J.M.A. acknowledges the CNPq Brazilian research agency for funding (grant No. 311589/2021-9).
B.L.C.M. (grant No. 305804/2022-7), J.R.M. (grant No. 308928/2019-9), and I.C.L. (grant No. 313103/2022-4) acknowledge CNPq research fellowships.
This work includes data collected by the TESS mission. The NASA Explorer Program provides funding for the TESS mission. All the TESS data used in this paper can be found in MAST:\dataset[10.17909/0kat-v379]{http://dx.doi.org/10.17909/0kat-v379}.

\end{acknowledgments}

\renewcommand{\thesection}{\Alph{section}}
\renewcommand{\thefigure}{\thesection\arabic{figure}}

\pagestyle{empty}
\nolinenumbers
\setcounter{section}{0}

\section*{Supplementary Material}

\section{Effect of data gaps and noise levels on period recovery in TESS-simulated light curves}
\setcounter{figure}{0}

\begin{figure}[h!]
	\includegraphics[width=3.5in]{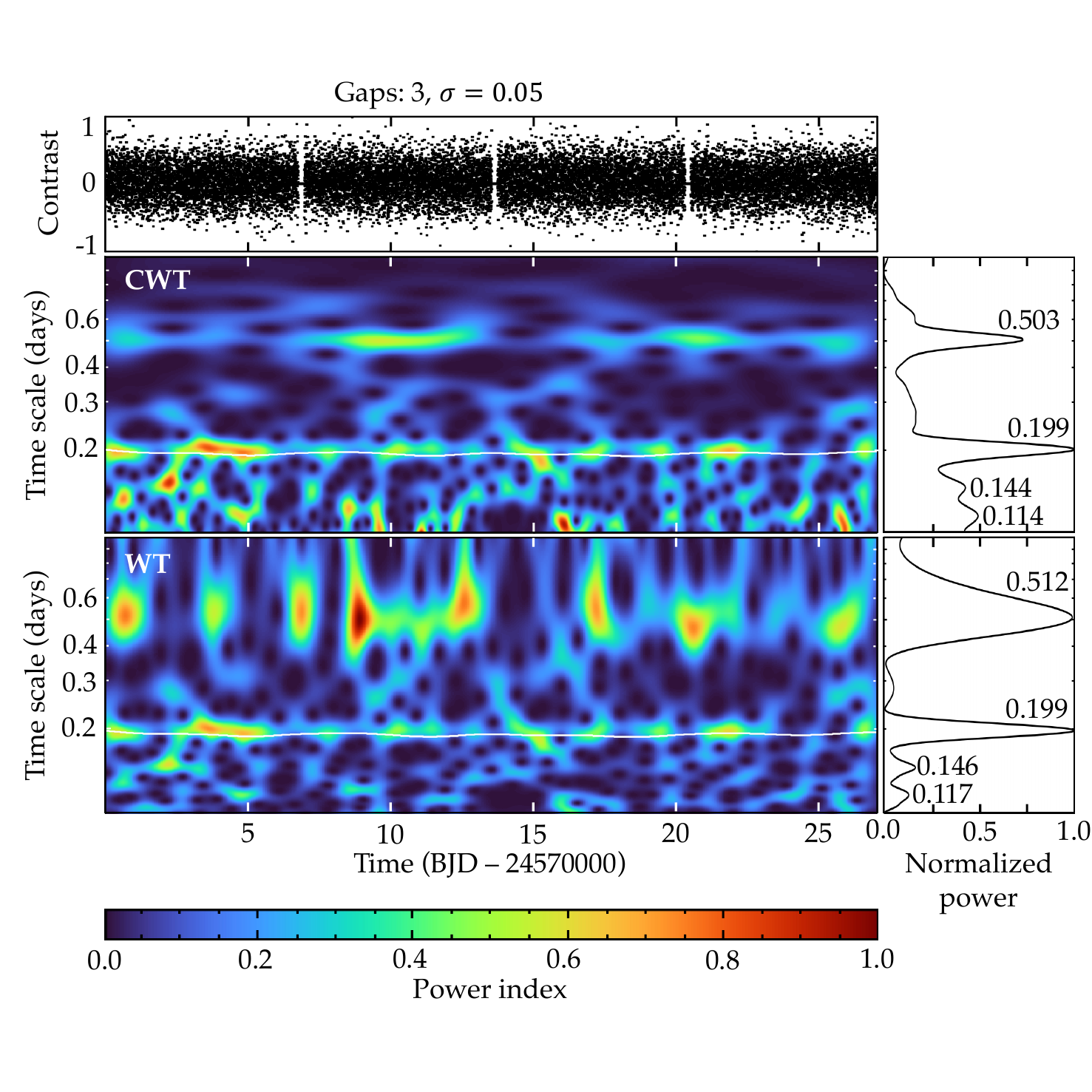}
	\includegraphics[width=3.5in]{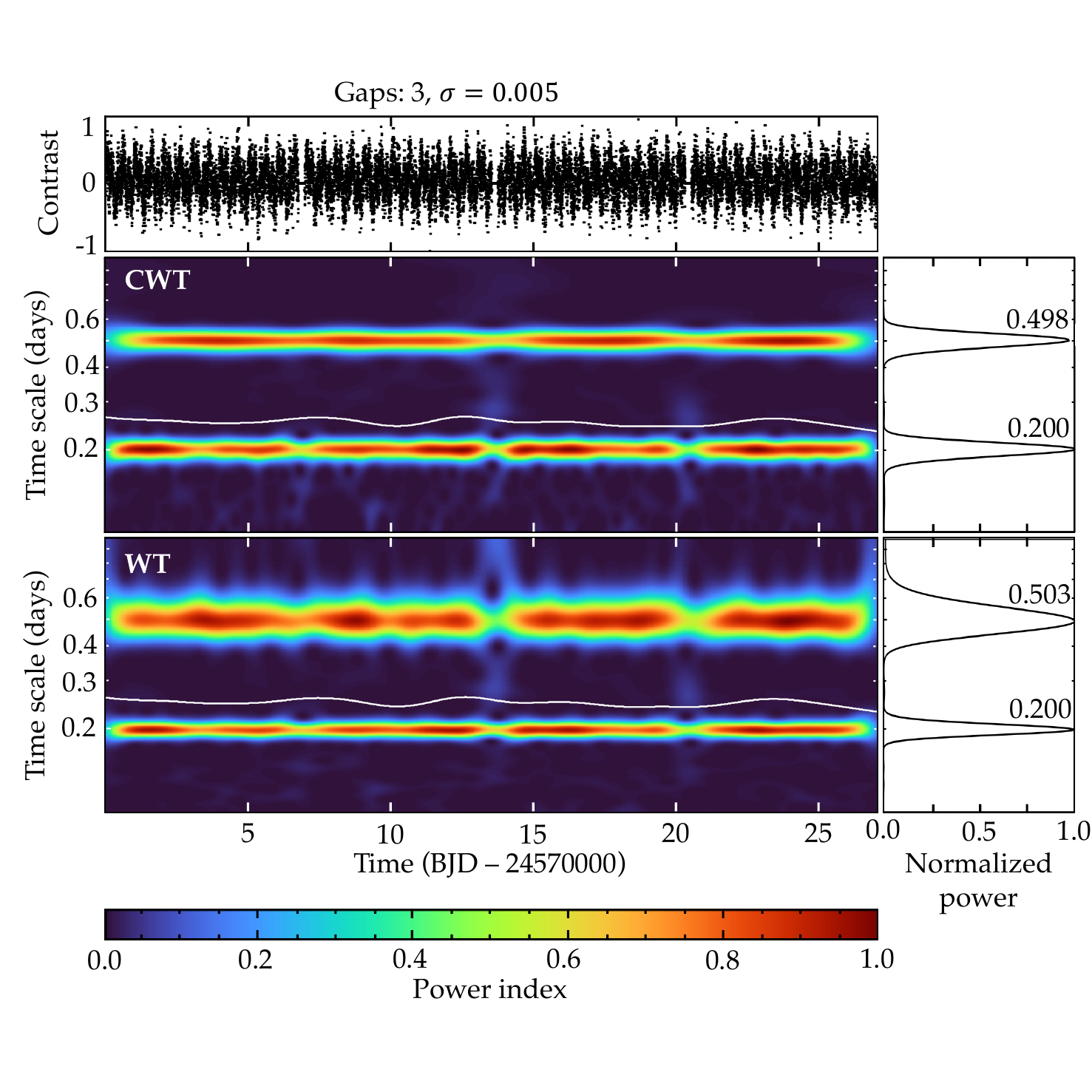}
    \begin{center}
    	\includegraphics[width=3.5in]{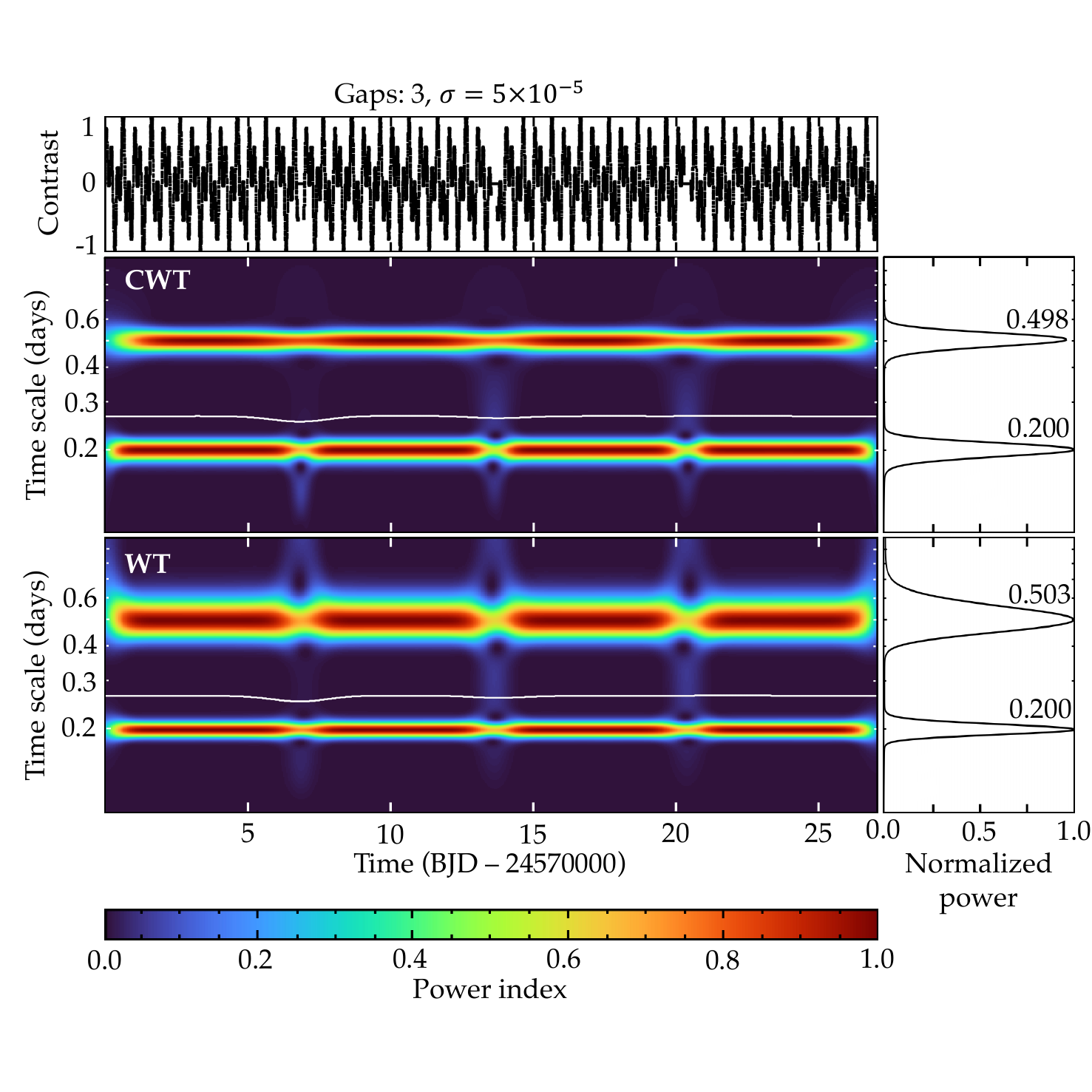}        
    \end{center}
\caption{Simulated TESS-like light curves with 3 gaps and fixed periods of 0.2 and 0.5 d. The top panels show the light curves, while the second and third rows display the CWT and WT local maps. The rightmost panels present the corresponding global spectra. Results are shown for three Gaussian noise levels: $\sigma$=0.05, 0.005, and 0.00005. The color palette represents the normalized power index levels, and the solid white lines indicate the dominant periods.\label{fig:Gaps3A}}
\end{figure}

\begin{figure}[h!]
	\includegraphics[width=3.5in]{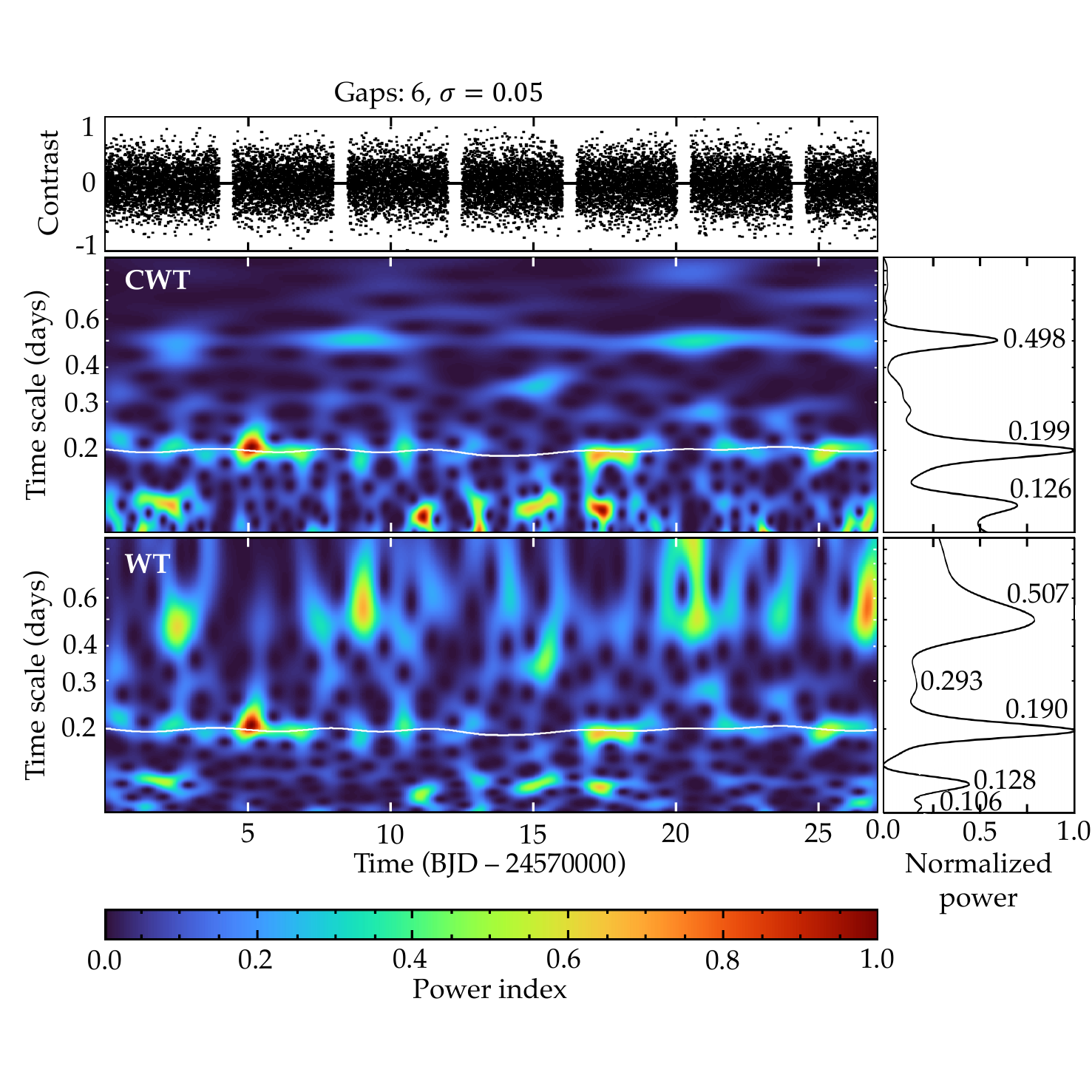}
	\includegraphics[width=3.5in]{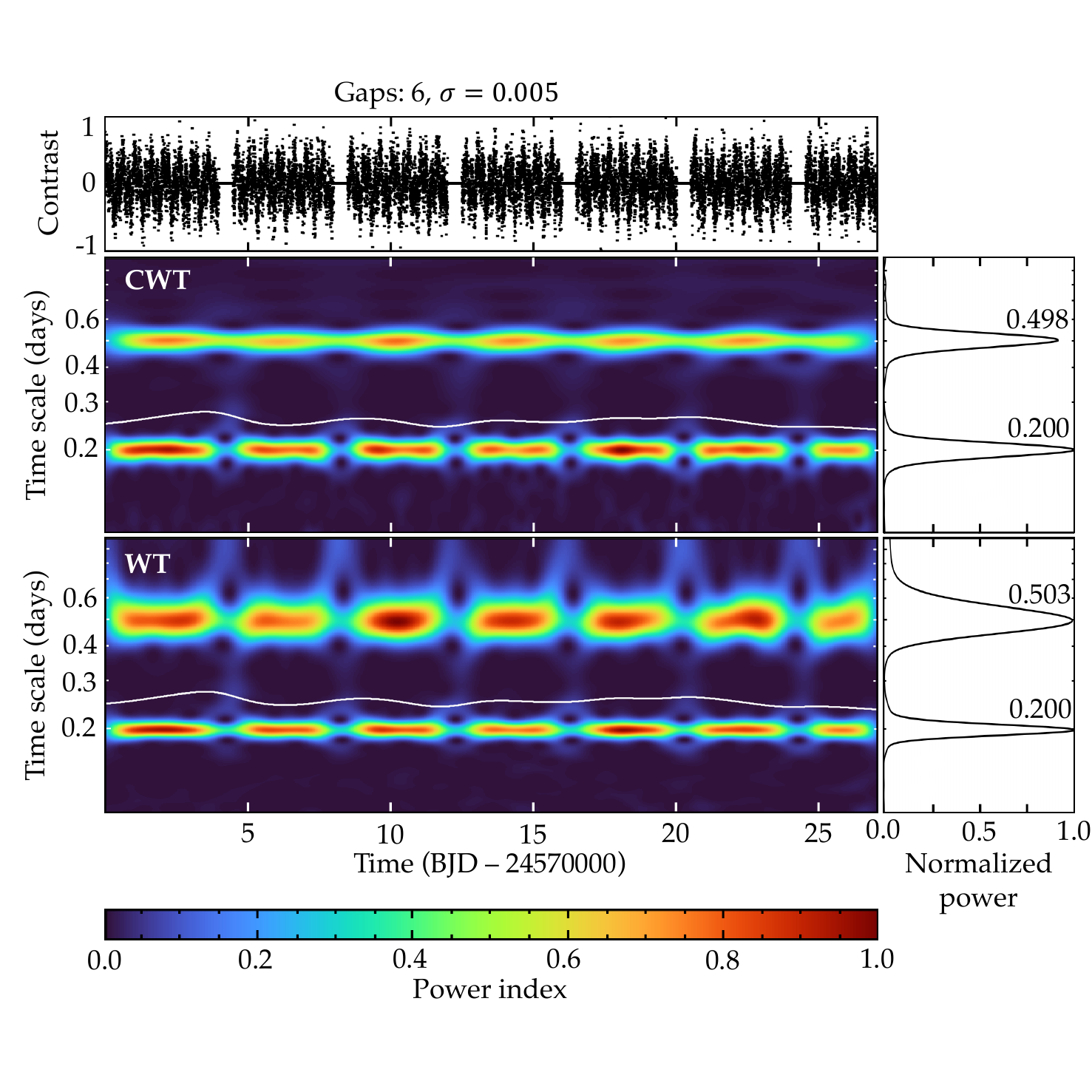}
    \begin{center}
    	\includegraphics[width=3.5in]{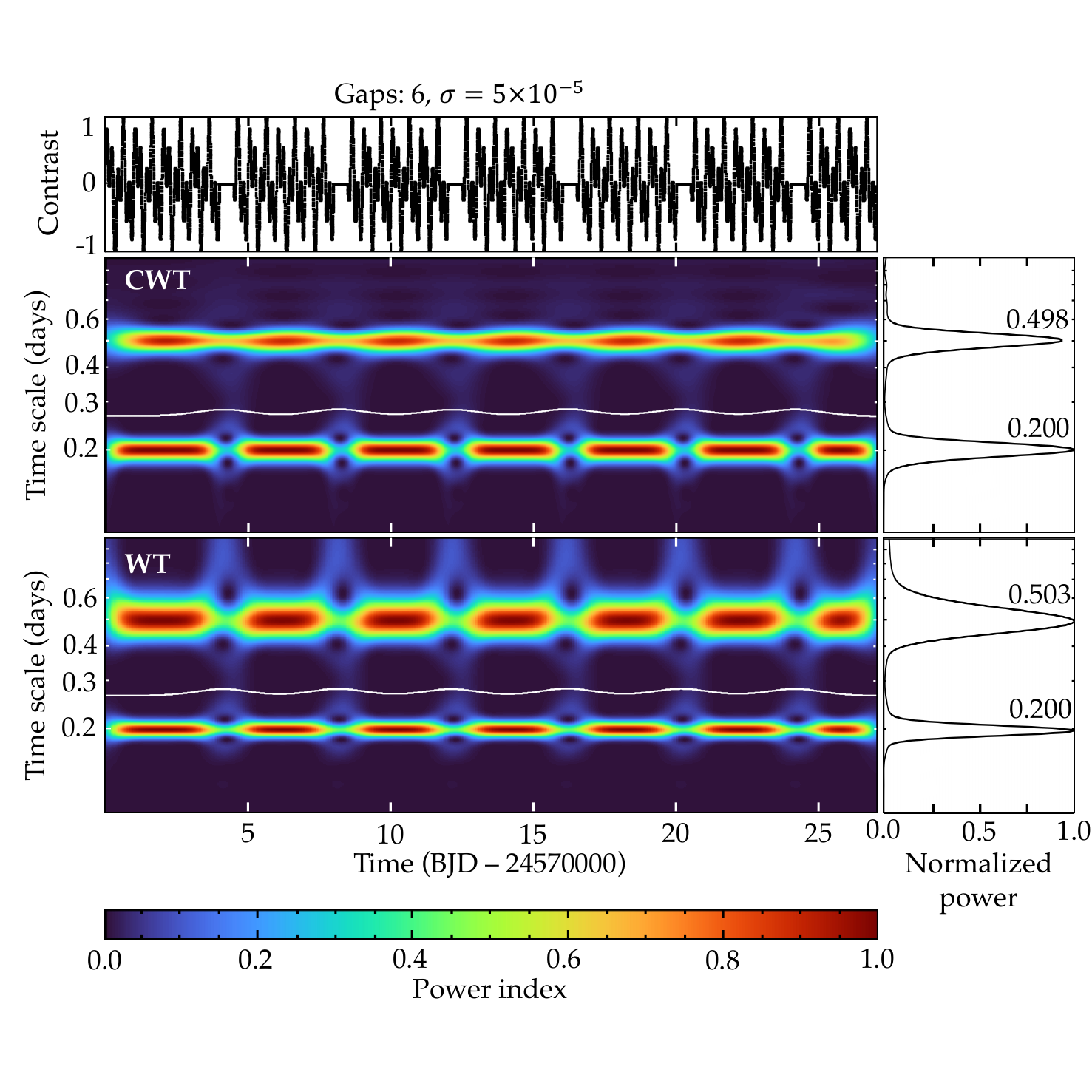}        
    \end{center}
\caption{Same as Figure \ref{fig:Gaps3A}, but for simulated light curves with 6 gaps.\label{fig:Gaps6A}}
\end{figure}

\begin{figure}[h!]
	\includegraphics[width=3.5in]{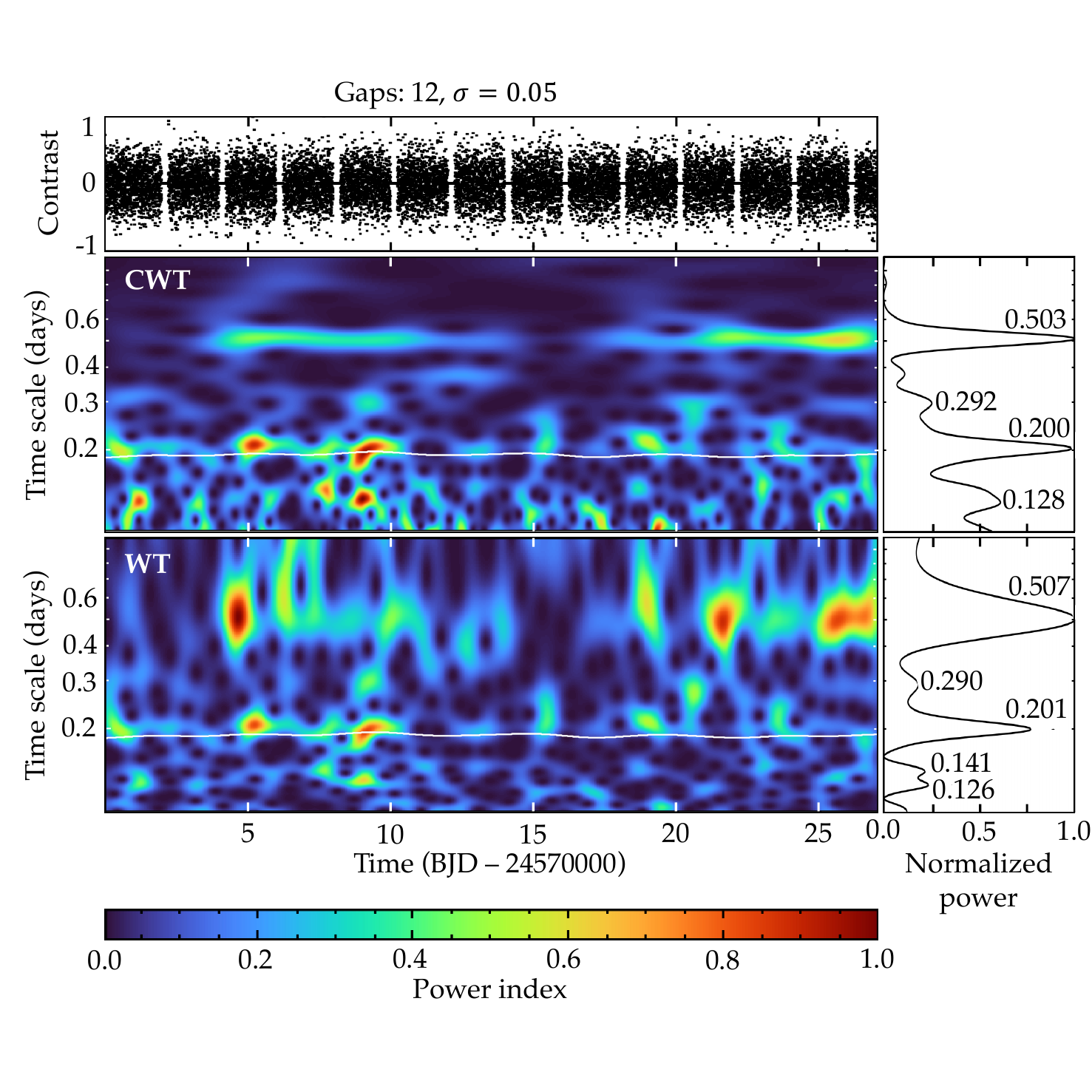}
	\includegraphics[width=3.5in]{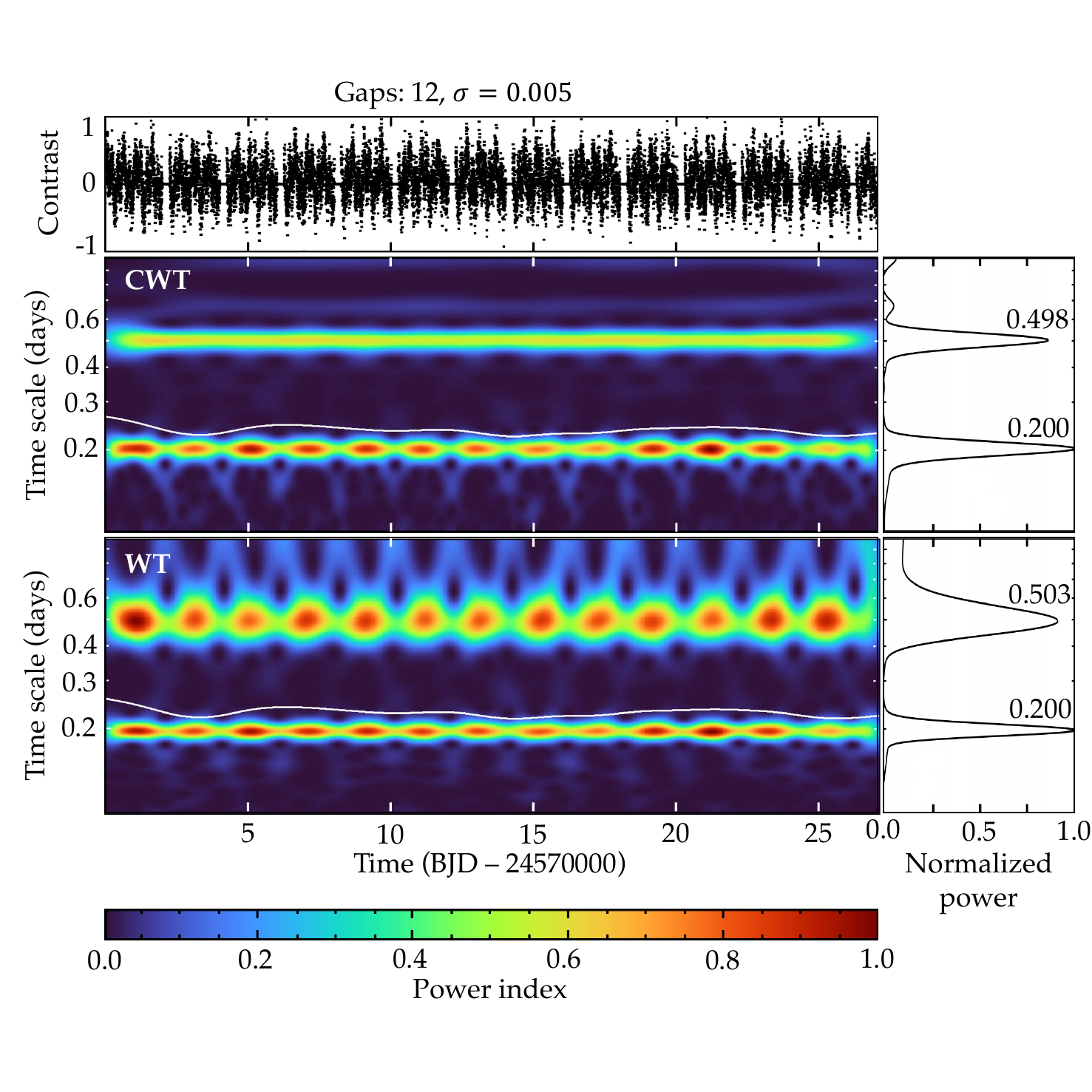}
    \begin{center}
    	\includegraphics[width=3.5in]{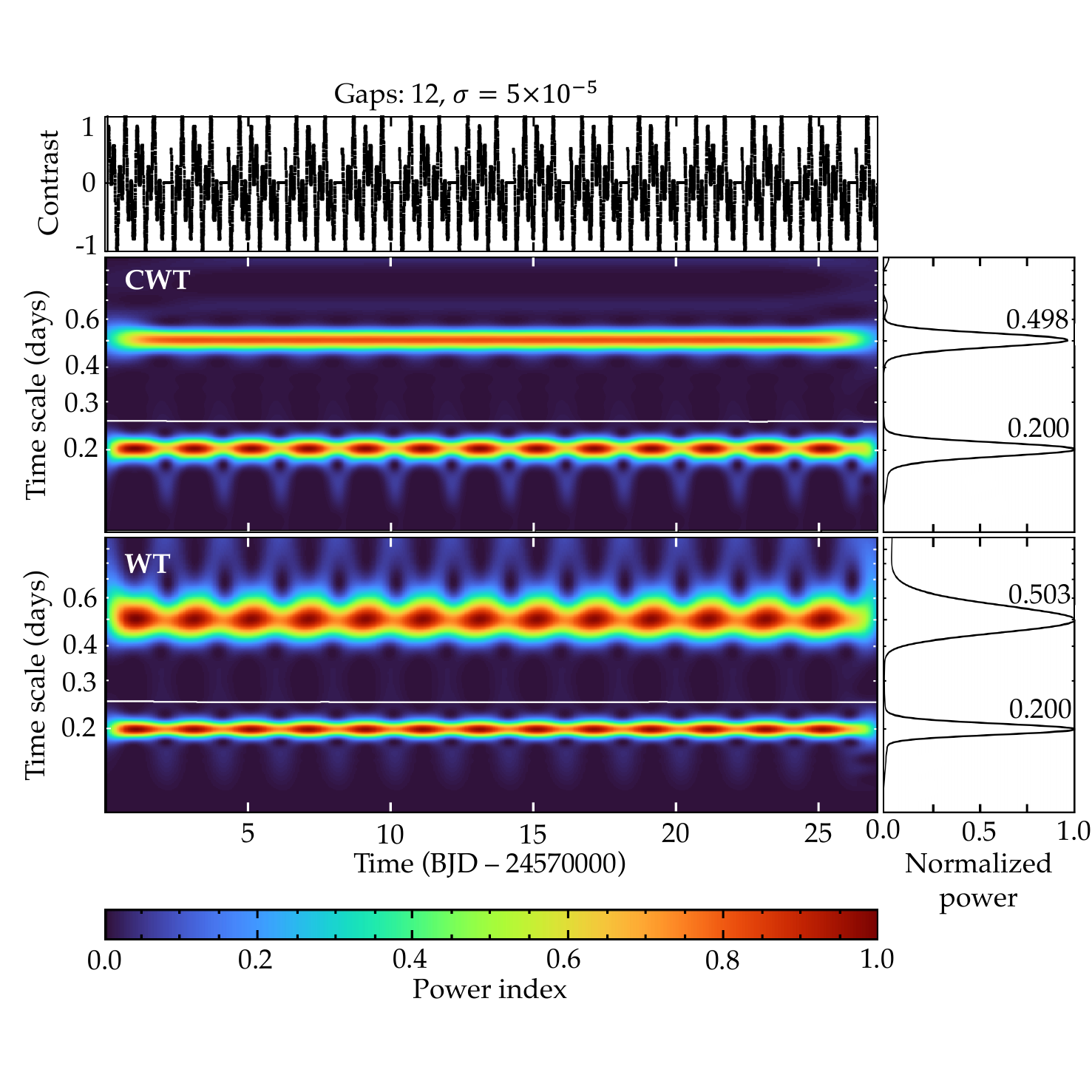}        
    \end{center}
\caption{Same as Figure \ref{fig:Gaps3A}, but for simulated light curves with 12 gaps.\label{fig:Gaps12A}}
\end{figure}

Figures \ref{fig:Gaps3A}, \ref{fig:Gaps6A}, and \ref{fig:Gaps12A} present TESS-simulated light curves with fixed periods of 0.2 and 0.5 d, including 3, 6, and 12 gaps, respectively. Each figure shows results for three Gaussian noise levels: $\sigma$=0.05, 0.005, and 0.00005. The panels illustrate the corresponding light curves, time-frequency maps, and global spectra obtained with WT and CWT, showing how both the number of gaps and the level of noise affect the recovery of the input periods. 

\clearpage
\newpage
\nolinenumbers
\thispagestyle{empty}

\section{Effect of different period values in TESS-simulated light curves}
\setcounter{figure}{0}

\begin{figure}[h!]
	\includegraphics[width=3.5in]{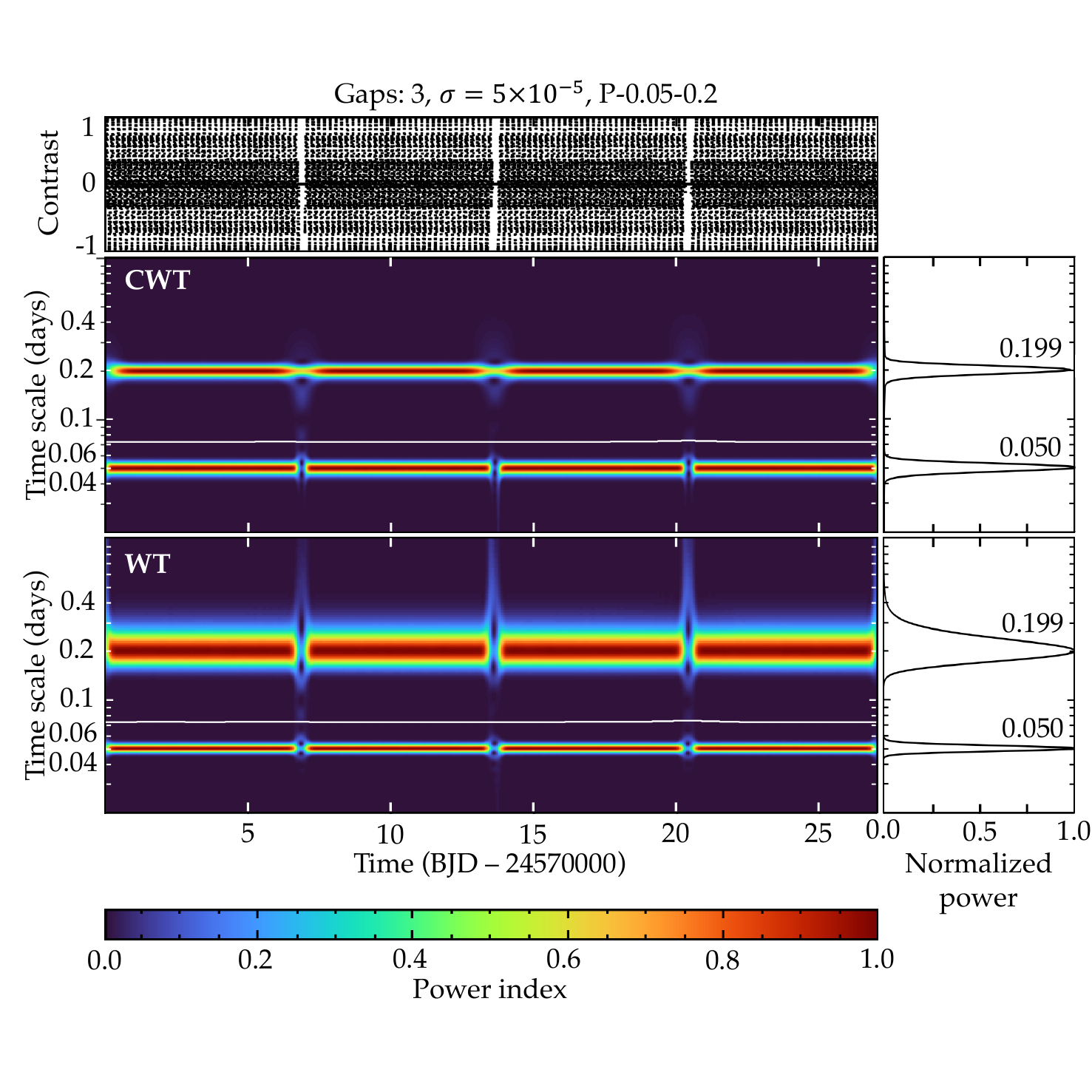}
	\includegraphics[width=3.5in]{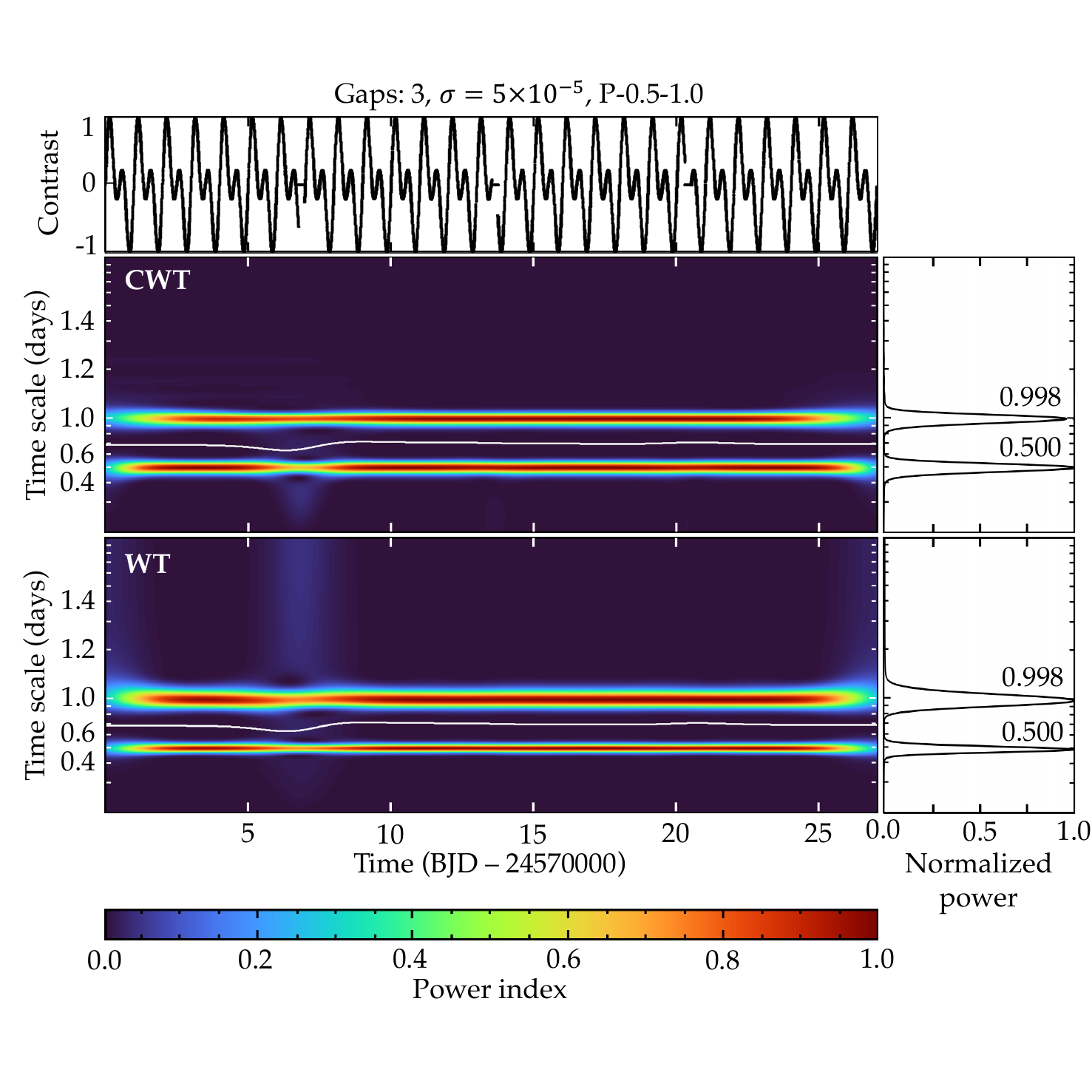}
    \begin{center}
    	\includegraphics[width=3.5in]{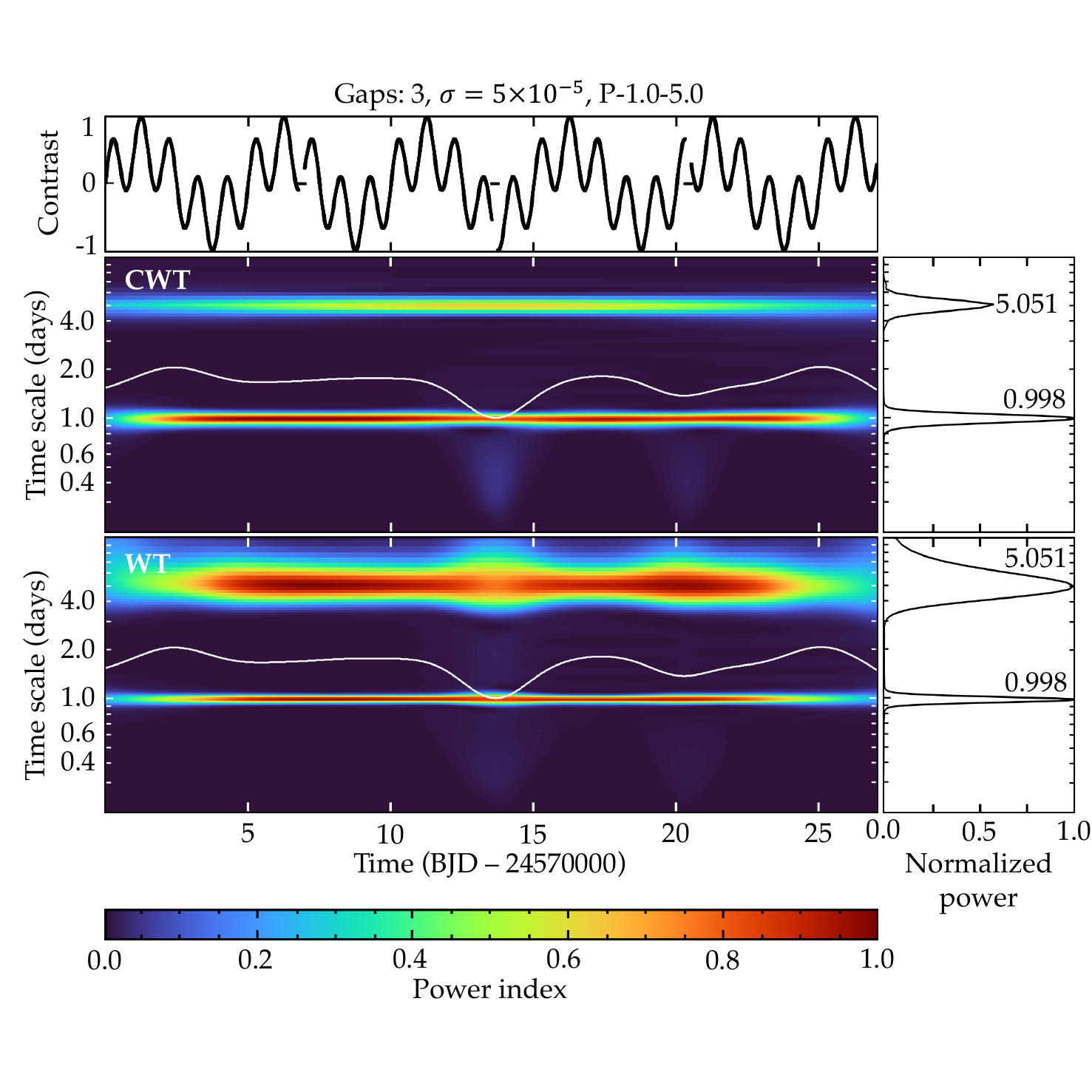}        
    \end{center}
\caption{Simulated TESS-like light curves with 3 gaps and fixed noise level $\sigma$=0.00005. The top panels show the light curves, while the second and third rows display the CWT and WT local maps. The rightmost panels present the corresponding global spectra. The three panels show tests for different period pairs: 0.05-0.2 d, 0.5-1.0 d, and 1.0-5.0 d. The color palette represents the normalized power index levels, and the solid white lines indicate the dominant periods.\label{fig:Gaps3B}}
\end{figure}

\begin{figure}[h!]
	\includegraphics[width=3.5in]{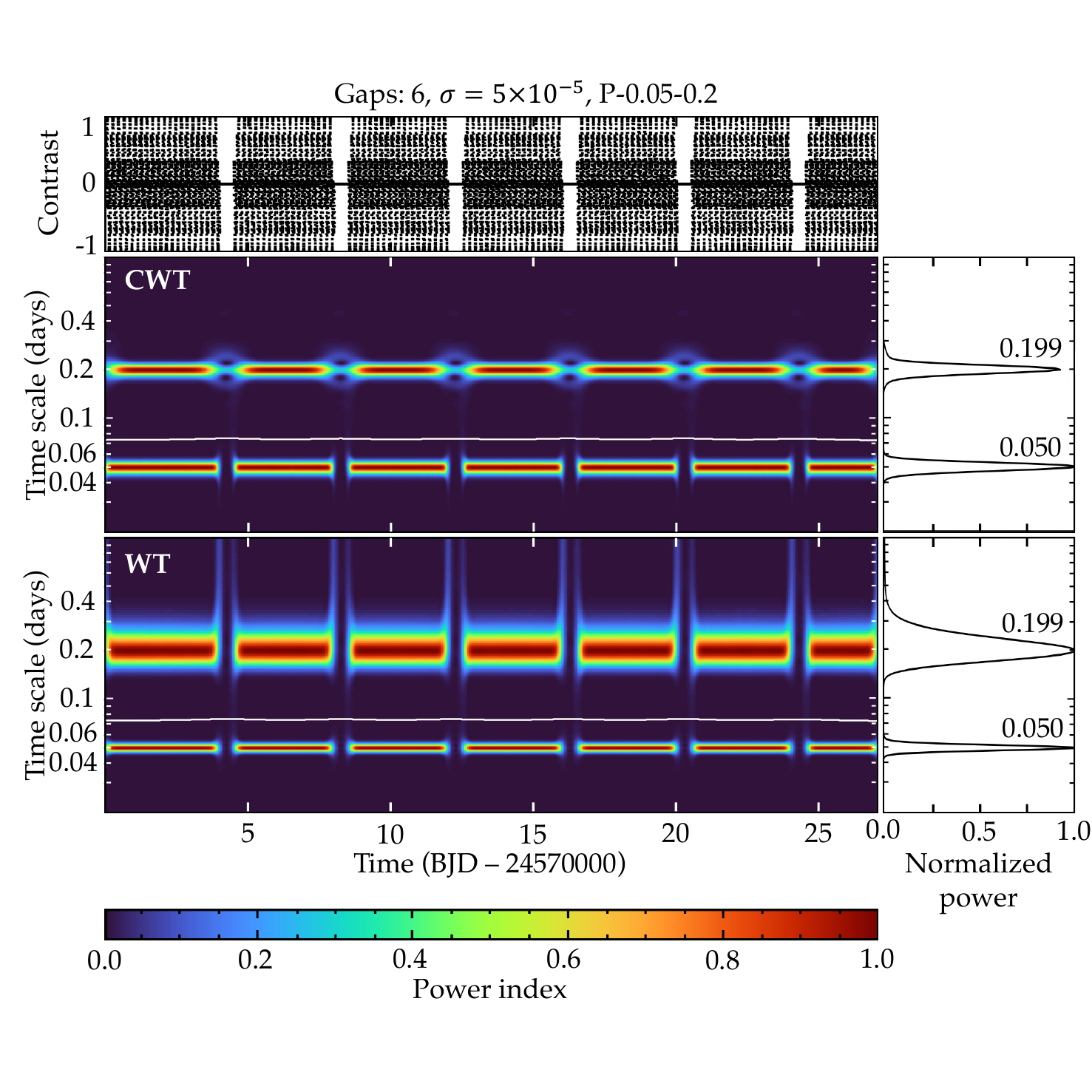}
	\includegraphics[width=3.5in]{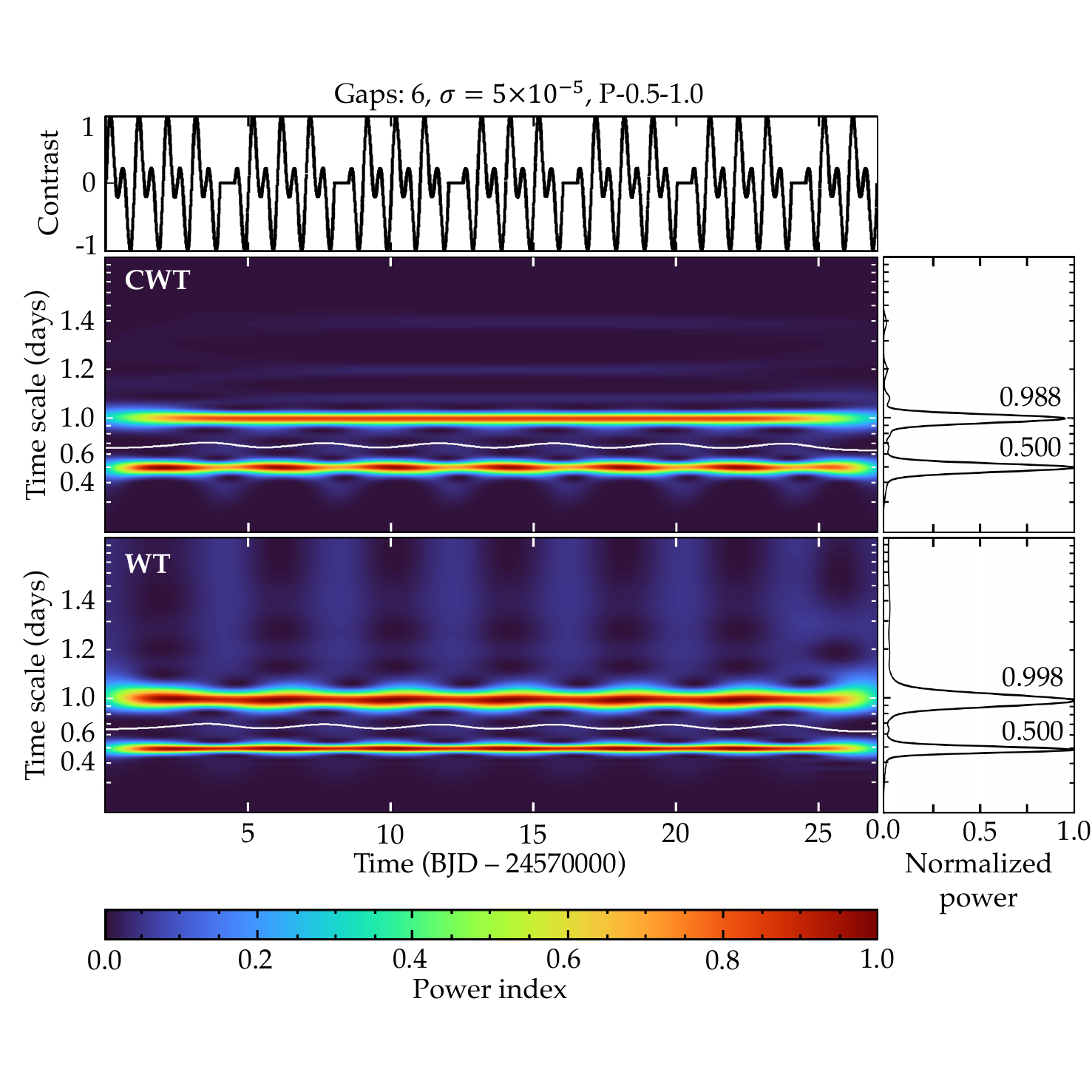}
    \begin{center}
    	\includegraphics[width=3.5in]{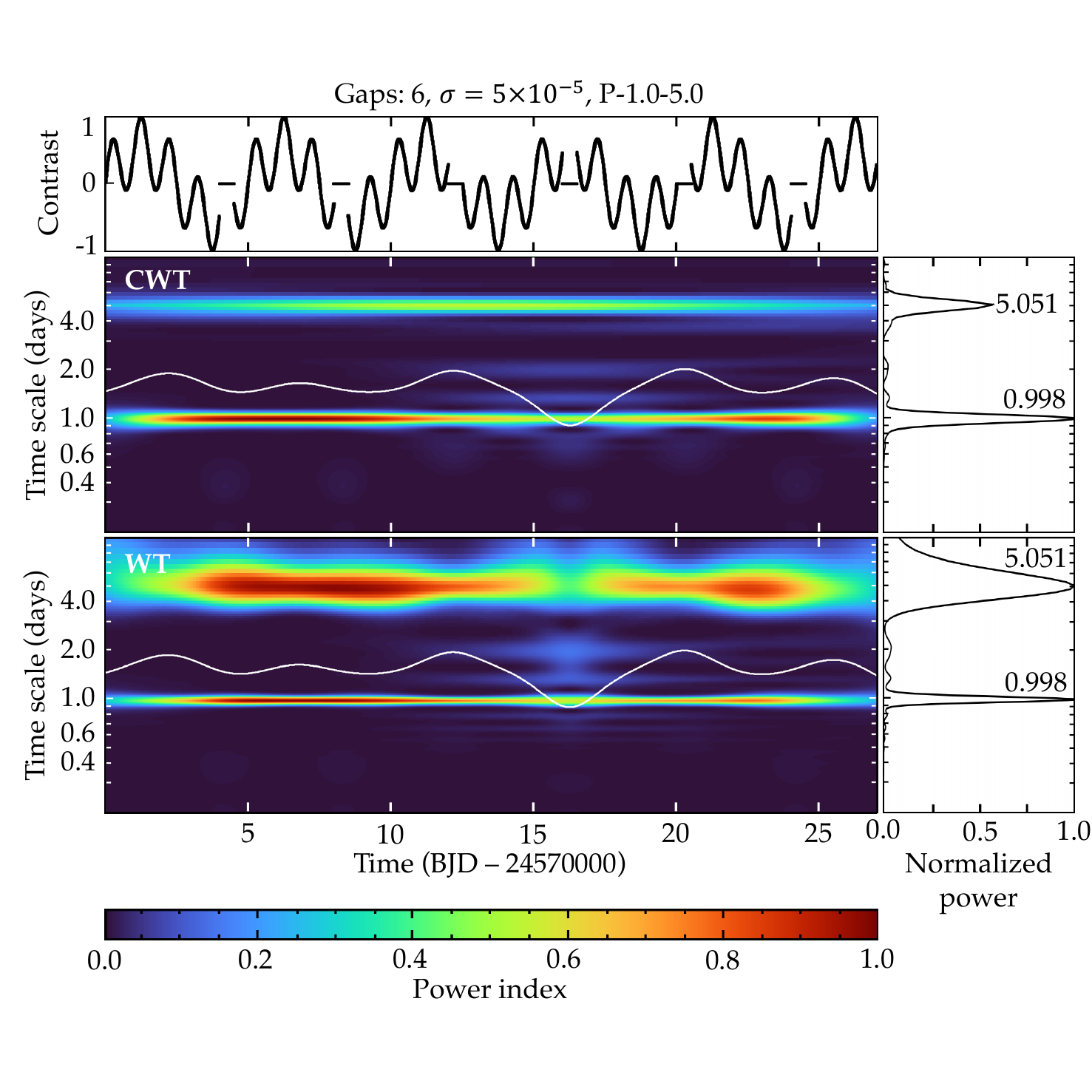}        
    \end{center}
\caption{Same as Figure \ref{fig:Gaps3B}, but for simulatesd light curves with 6 gaps.\label{fig:Gaps6B}}
\end{figure}

\begin{figure}[h!]
	\includegraphics[width=3.5in]{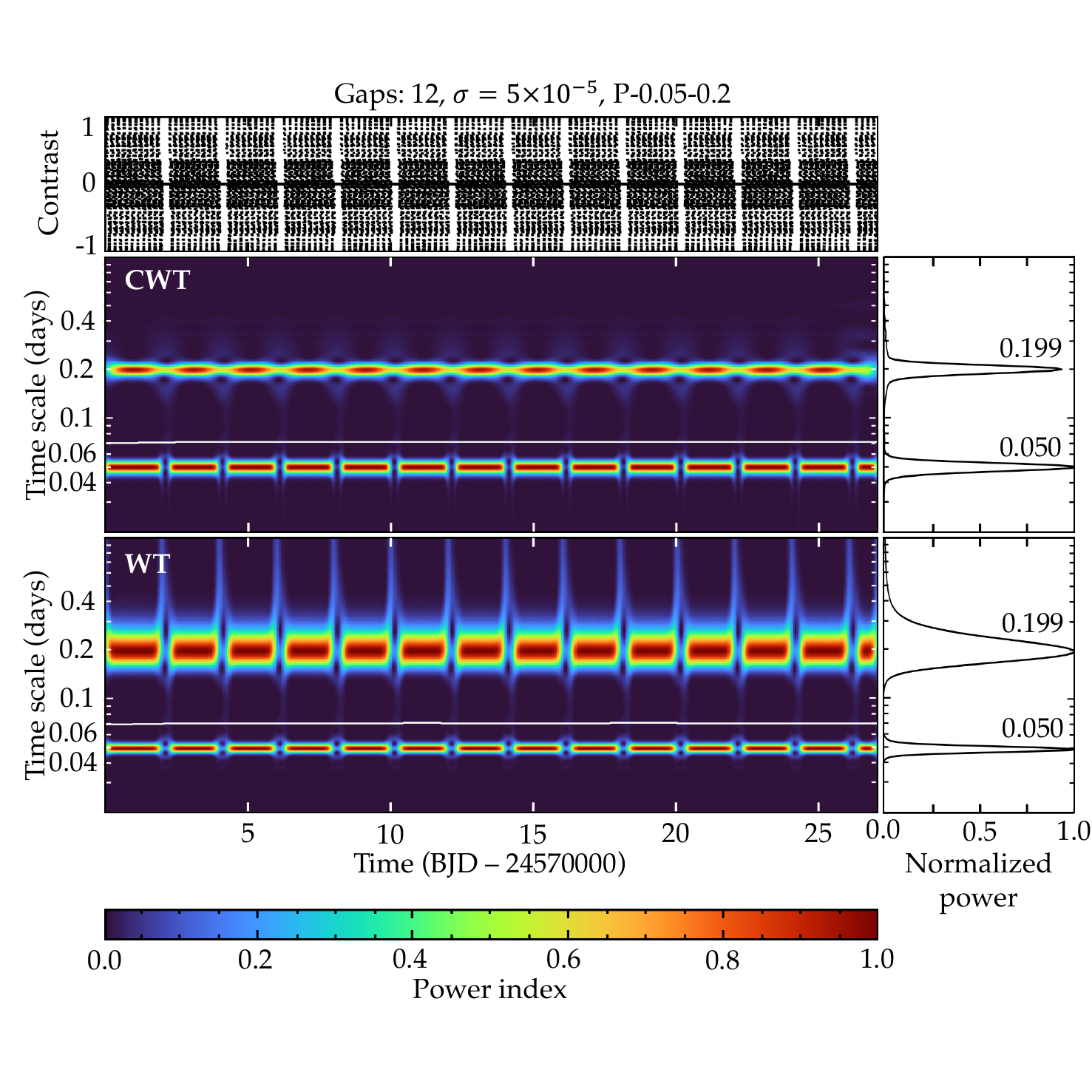}
	\includegraphics[width=3.5in]{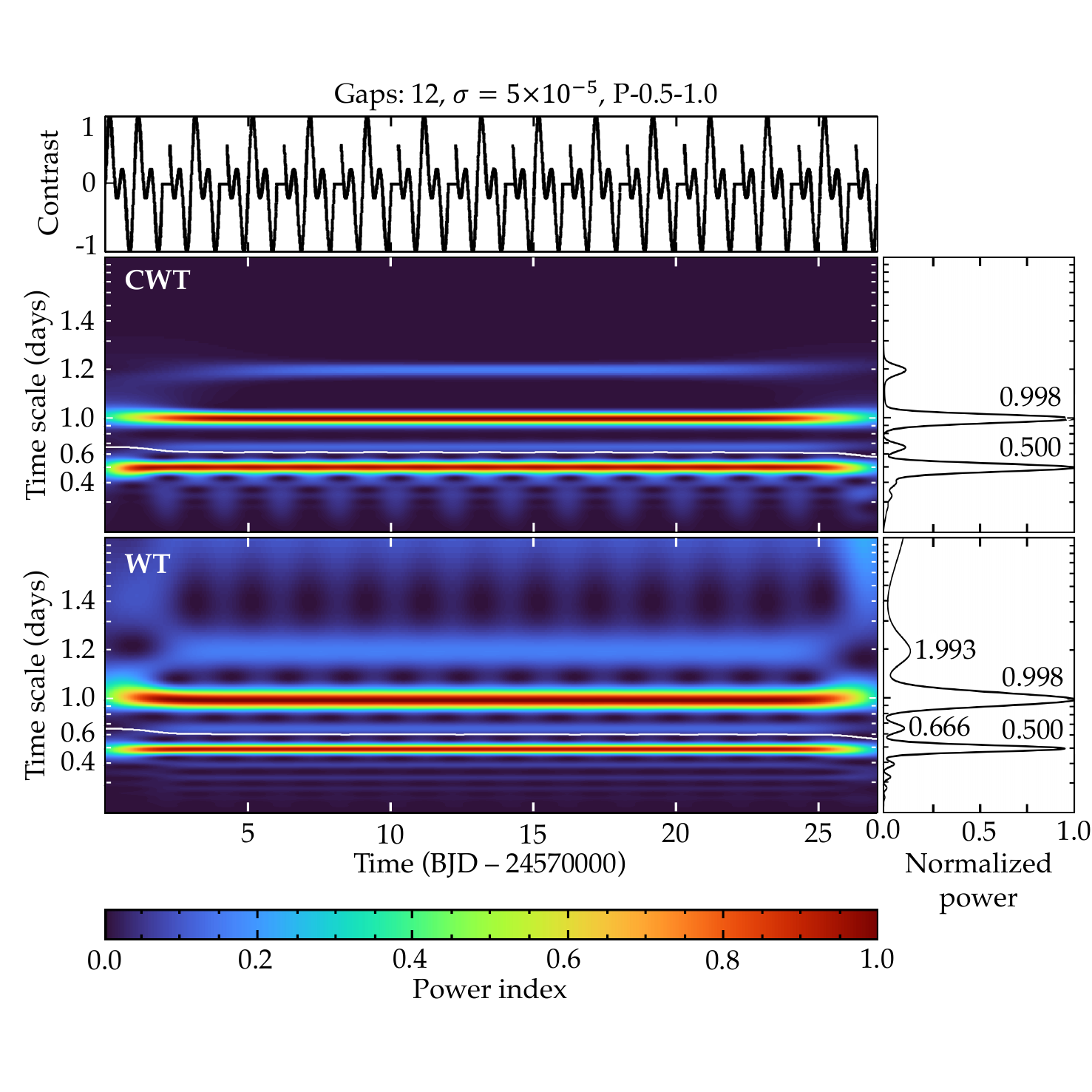}
    \begin{center}
    	\includegraphics[width=3.5in]{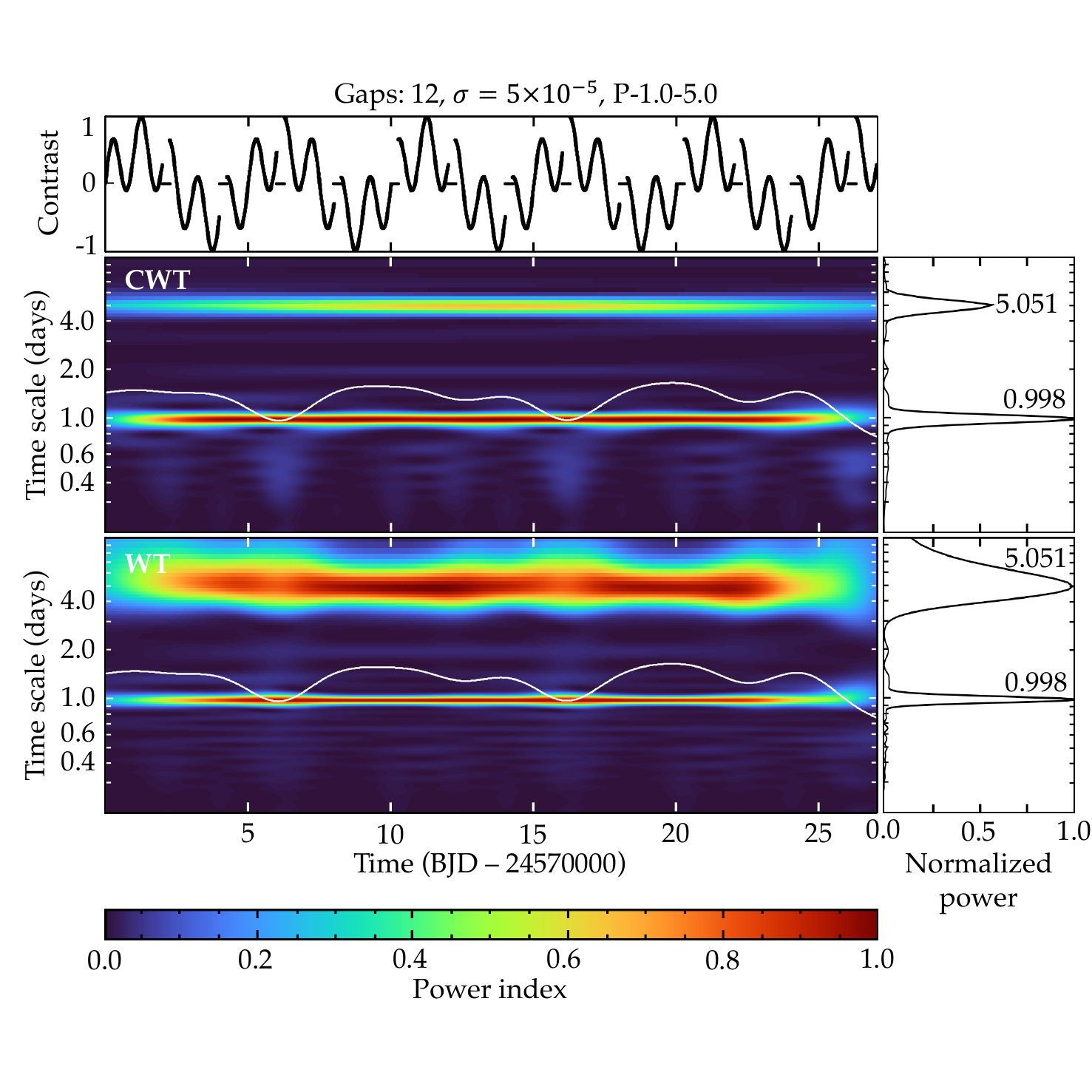}        
    \end{center}
\caption{Same as Figure \ref{fig:Gaps3B}, but for simulated light curves with 12 gaps.\label{fig:Gaps12B}}
\end{figure}

Figures \ref{fig:Gaps3B}, \ref{fig:Gaps6B}, and \ref{fig:Gaps12B} present TESS-simulated light curves with the lowest noise level fixed ($\sigma$=0.00005) and with 3, 6, and 12 gaps, respectively. Each figure shows results for three different period pairs: 0.05-0.2 d, 0.5-1.0 d, and 1.0-5.0 d. 

\end{document}